\documentclass[manuscript]{aastex}

\newcommand{\photu}{ph\ cm$^{-2}$ s$^{-1}$ sr$^{-1}$ \AA$^{-1}$}
\newcommand{\galex}{{\it GALEX}}

\shorttitle{GALEX Diffuse Observations of the Sky: The Data}
\shortauthors{Murthy}

\begin{document}

\title{GALEX Diffuse Observations of the Sky: The Data}

\author{Jayant Murthy}
\affil{Indian Institute of Astrophysics, Bangalore 560 034}
\email{jmurthy@yahoo.com}

\begin{abstract}

I present tabulations of the diffuse observations made by the \galex\ spacecraft in two UV bands (FUV: 1539 \AA\ and NUV: 2316 \AA) from the (almost) final data release of the \galex\ spacecraft (GR6/GR7). This data release includes all the FUV observations and the majority of the NUV observations. I discuss overall trends in the data but the primary purpose is to make the data available to the public. These data files described in this paper are hosted by the Mikulski Archive for Space Telescopes (MAST) at the Space Telescope Science Insitutute from whence they may be downloaded. For ease of use, I have also created maps of the diffuse radiation in both bands over the entire observed sky at 6\arcmin\ resolution.

\end{abstract}


\section{Introduction}

The study of the diffuse ultraviolet (UV) background was pioneered through rocket observations by \citet{Hayakawa1969} who observed radiation in the direction of the Galactic anti-center. They found that the intensity of this radiation could be approximated by  a plane-parallel model for the interstellar dust distribution. \citet{Witt1973}  first observed the diffuse radiation from a spacecraft using a UV photometer aboard the Orbiting Astronomical Observatory (OAO-2) and modeled it using  strongly forward scattering interstellar dust grains with an albedo (a) of about 0.5 and a phase function asymmetry factor (g) of about 0.75. These were followed by a series of rocket and spacecraft observations (reviewed by \citet{Bowyer1991, Henry1991, Murthy2009}) which obtained results that, at the time, seemed divergent. As the instrumentation improved, the data became more reliable and there was general agreement that the diffuse light was correlated with the amount of dust with a baseline level of 200 - 300 \photu\ at the Galactic poles, either from residual Galactic emission or from extragalactic light \citep{Bowyer1991, Henry1991}.

A major advance came with the launch of the {\it SPEAR} ({\it Spectroscopy of Plasma Evolution from Astrophysical Radiation}) mission \citep{Edelstein2006, Seon2011} which performed a spectral survey of 80\% of the sky in the wavelength range from 1370 -- 1710 \AA. This was followed by maps of the sky made using data from the {\it Galaxy Evolution Explorer} (\galex) by \citet{Murthy2010} at 30\arcmin\ resolution and \citet{Hamden2013} at 11\arcmin\ resolution. All three studies found good correlations between the UV emission and other tracers of dust and gas --- including the 100 \micron\ emission from interstellar dust observed by the {\it IRAS} ({\it Infrared Astronomy Satellite}) mission \citep{Schlegel1998}; 21 cm emission \citep{Kalberla2005}; and H $\alpha$ emission \citep{Finkbeiner2003} --- but with considerable scatter. In addition to the large scale structure associated with the Galactic dust distribution, there are smaller features such as the halos found around hot stars \citep{Murthy2011, Choi2013}.

Now that \galex\ has completed its mission and is no longer taking observations, I have extracted and compiled the diffuse background in both the far ultraviolet (FUV) and the near ultraviolet (NUV) channels at a resolution of 2\arcmin. The FUV data are complete while the NUV data are complete except for a small number of proprietary observations taken near the end of the mission. I present the data processing and the actual data in this work leaving a detailed analysis to further studies. An analogous catalog of the point sources observed by \galex\ has been published by \citet{BCS2014} and an overview of the science and data products from the mission has been given by \citet{Bianchi2014}. Although the primary resource for the diffuse radiation is the data files containing the diffuse background from each visit, I have also created Aitoff maps of the sky at 6\arcmin\ resolution for both bands.

\section{Data}

\begin{figure*}[t]
\includegraphics[width=3in]{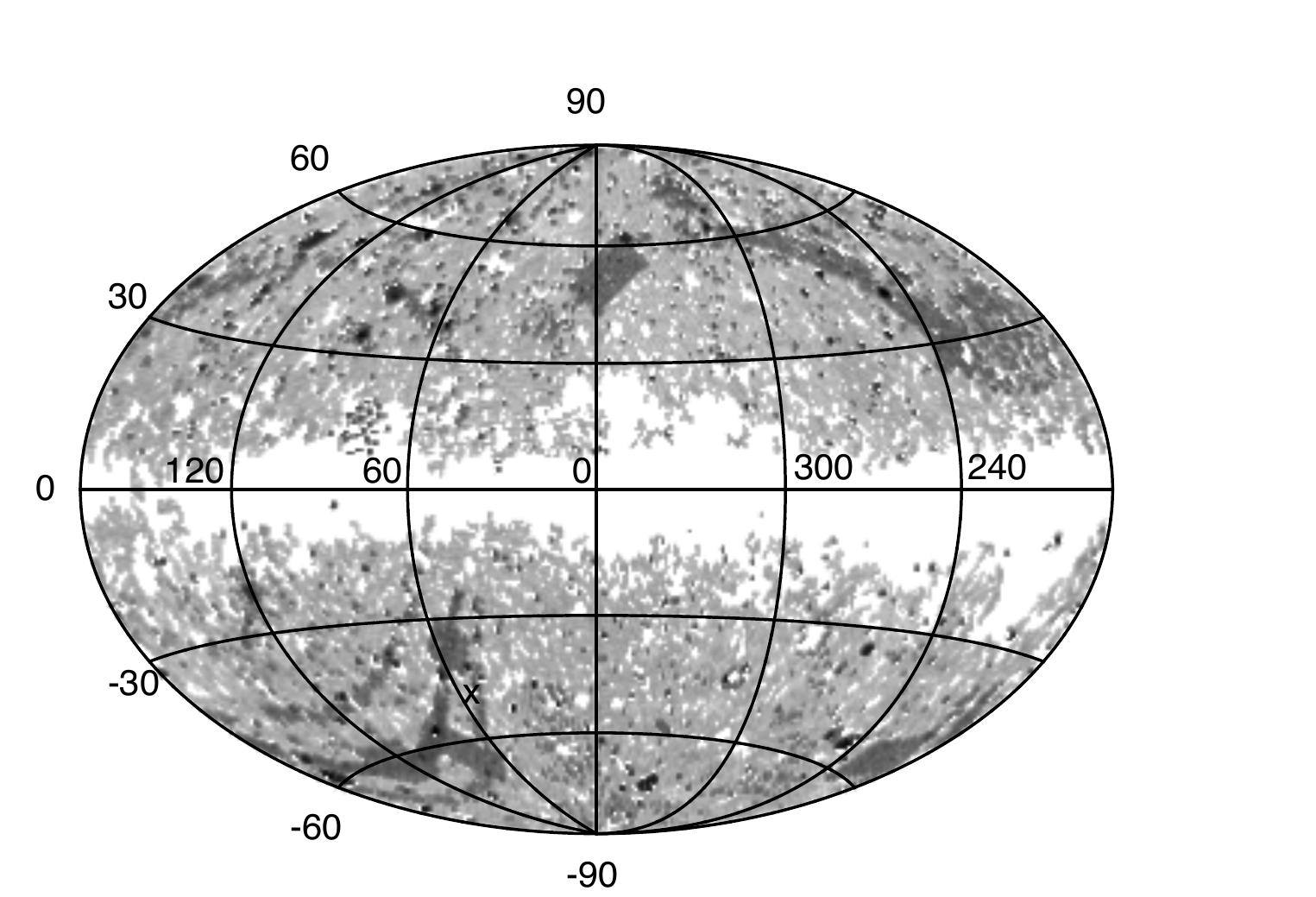}
\includegraphics[width=3in]{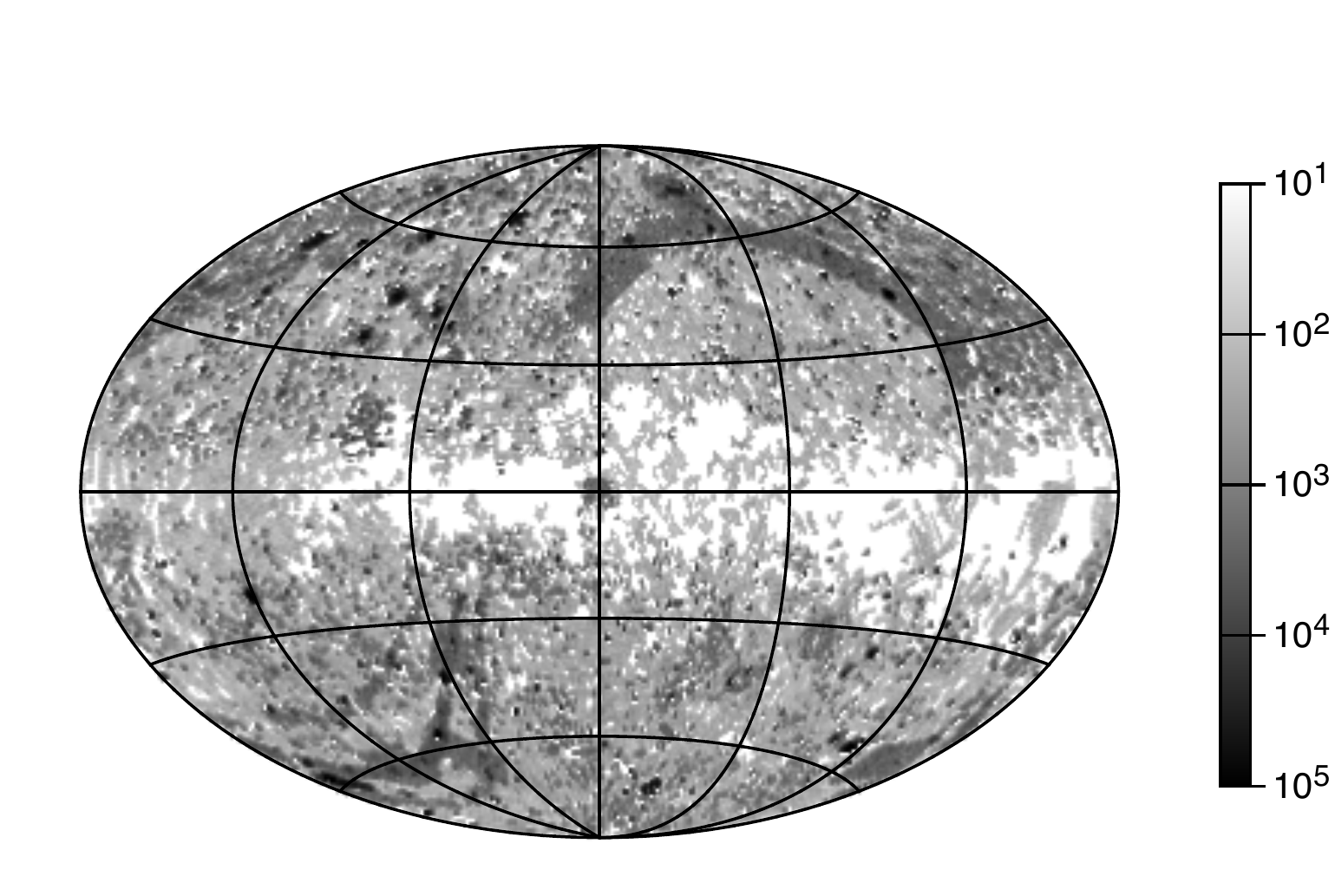}
\caption{Exposure time in seconds per pixel in the FUV (a) and NUV (b). The color bar shows the exposure time per pixel in seconds. White areas were not observed. Both plots are shown in Aitoff coordinates with the Galactic center in the center of the plot.}
\label{fig:exptime}
\end{figure*}

\begin{figure*}[t]
\includegraphics[width=3in]{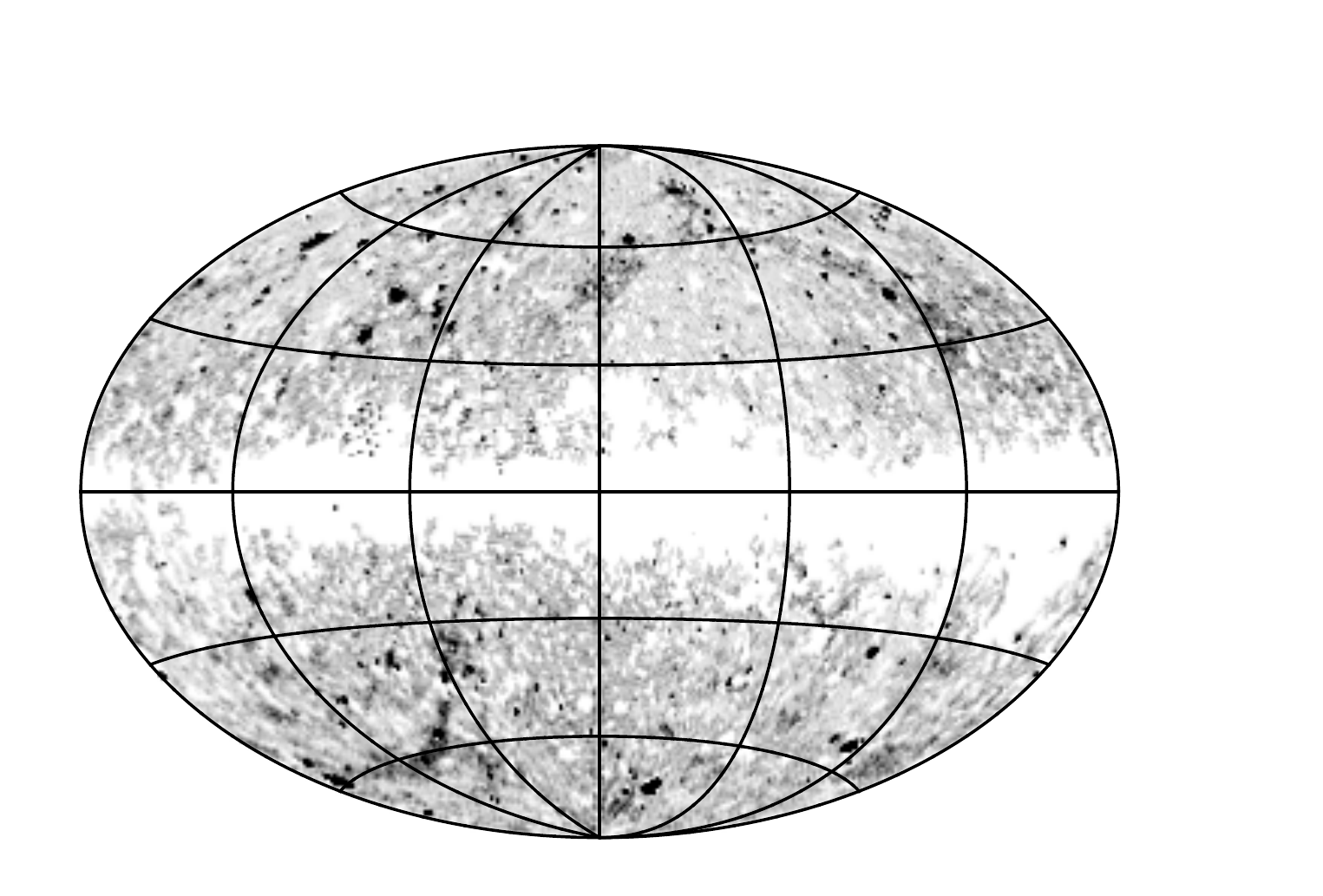}
\includegraphics[width=3in]{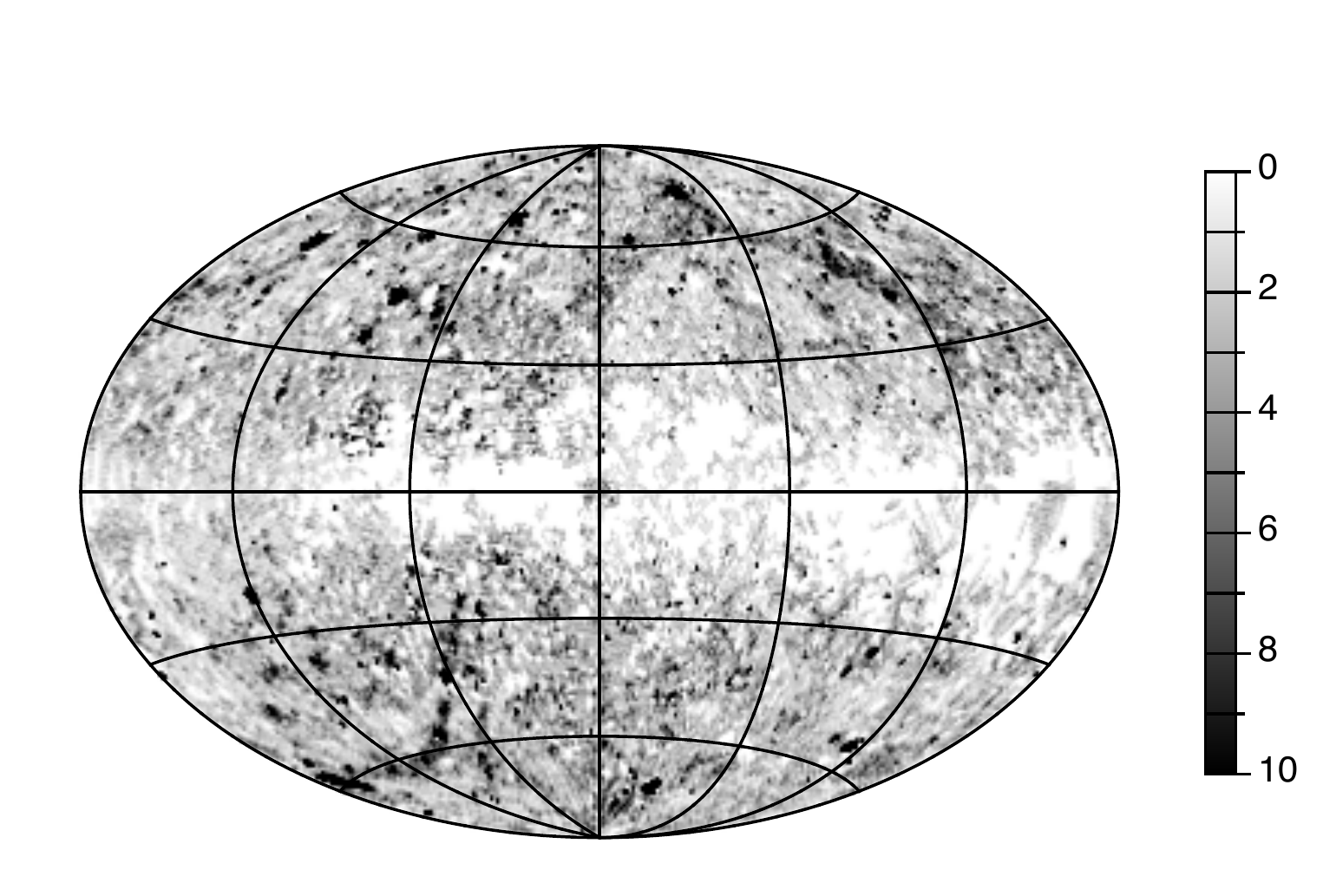}
\caption{Number of different visits for each pixel in the FUV (a) and NUV (b). The color bar shows the number of visits for each pixel. White areas were not observed. Galactic coordinates are as in Fig. \ref{fig:exptime}.}
\label{fig:visits}
\end{figure*}

\galex\ was launched in 2003 by a Pegasus rocket into an equatorial orbit at an altitude of about 600 km \citep{Martin2005}. The instrument used a 50 cm Ritchey-Chr\'{e}tien telescope to image an 0.6\degr\ radius field onto two detectors (FUV: 1344 -- 1786 \AA\ and NUV: 1771 -- 2831 \AA) with a spatial resolution of 5 - 10\arcsec . There were a total of 44,843 public observations dating from June 7, 2003 to June 2, 2012 in the GR6/GR7 data release. Observations continued until the spacecraft was decommissioned on June 28, 2013 and these data will be made public in due course. I have used only those data from the GR6/GR7 release in this work and have plotted the total exposure time over the sky for each of the FUV and the NUV channels in Fig.  \ref{fig:exptime}.   The FUV power supply had failed in May 2009 after intermittent problems and thus could not observe the Galactic plane and other high intensity regions. Because \galex\ was in a low Earth orbit (LEO), longer observations were split into individual exposures (visits) of no longer than about 1000 seconds and there were a total of 100,864 (actually 100,865 but one visit was unreadable from the archives) such visits (Fig. \ref{fig:visits}). Note that a given point in the sky may have been observed multiple times either through completely different observations or through a single observation with multiple visits. These visits were used to differentiate the terrestrial and Solar System components of the diffuse background from the astrophysical components \citep{Murthy2014}.

The \galex\ pipeline, data products, and the instrumental calibration have been described by \citet{Morrissey2007} but are intended for the scientific analysis of point sources and extended objects. I have used the standard products and further processed them to extract the diffuse background as described by \citet{Sujatha2010}. I started with the image files ({\it fd-int} and {\it nd-int}) from the GR6/GR7 release, which includes all imaging observations made by \galex, except for a few proprietary observations taken as part of the \galex\ Complete the All-sky UV Survey Extension (CAUSE) program. Each observation included a merged catalog file ({\it xd-mcat}) containing all point sources detected in the field in either of the two detectors with their maximum extent in x and y (\textit{xmin\_image, ymin\_image, xmax\_image, ymax\_image}) and I masked this region around each source in the two image files. Although the wings of the stellar point spread function may extend far from the star, they are at a level of less than 2\% of the stellar flux \citep{Morrissey2007} and will not contribute significantly to the diffuse flux, particularly as the contribution from stars to the total is less than 10\% in the FUV and 30\% in the NUV in the majority of locations. The stellar flux in each field was calculated from the source magnitudes tabulated in the merged catalog. I then binned the remaining pixels such that each output pixel is comprised of 80 input pixels in each direction with an effective resolution of 2\arcmin\ per binned pixel. Note that I simply ignored all masked pixels in the binning, effectively replacing them with the average over the entire square.

It is interesting to consider the statistics in the \galex\ data for a representative diffuse signal of 100 \photu\ in a typical AIS observation. Assuming a spectrally flat signal and integrating with the \galex\ calibration curve\footnote{\url{http://galexgi.gsfc.nasa.gov/docs/galex/Documents/PostLaunchResponseCurveData.html}}, I obtain a total count rate of 350 counts s$^{-1}$ over the entire FOV or 35,000 counts over the 100 second AIS exposure time implying an error of 187 counts over the detector surface (0.5 \photu). This is, of course, much higher if we use a smaller area and the uncertainty due to photon noise is 18 \photu\ in the FUV and 8 \photu\ in the NUV for the 2\arcmin\ pixel that we use. The deviations in a typical \galex\ field are much greater than this suggesting that the diffuse background has structure on scales of 2\arcmin\ or less.  For comparison, the uncertainty in the diffuse background per 1.5\arcsec\ \galex\ pixel is 1200 \photu\ in the FUV and 650 \photu\ in the NUV. 

\begin{figure*}[t]
\includegraphics[width=3in]{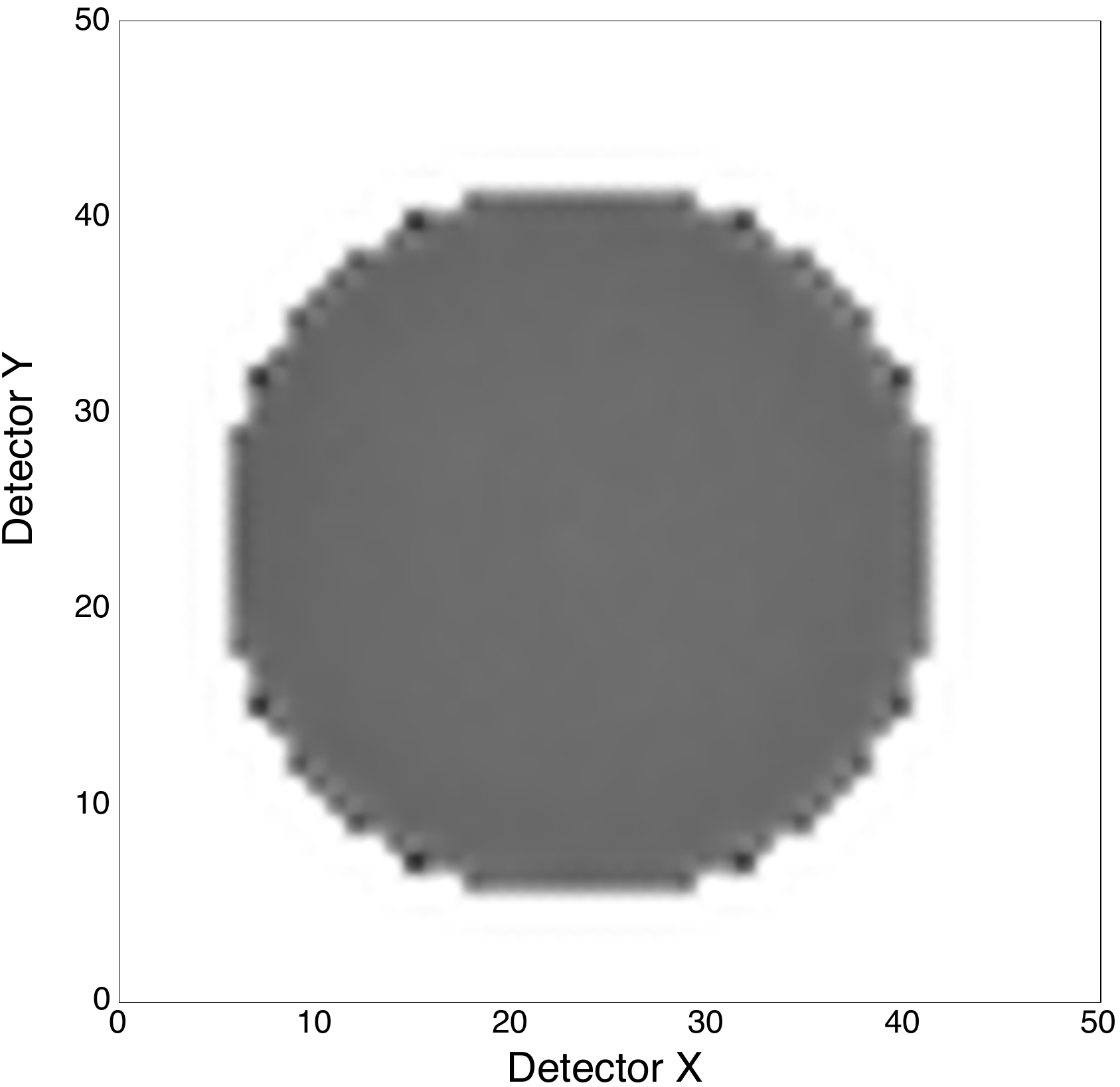}
\includegraphics[width=3in]{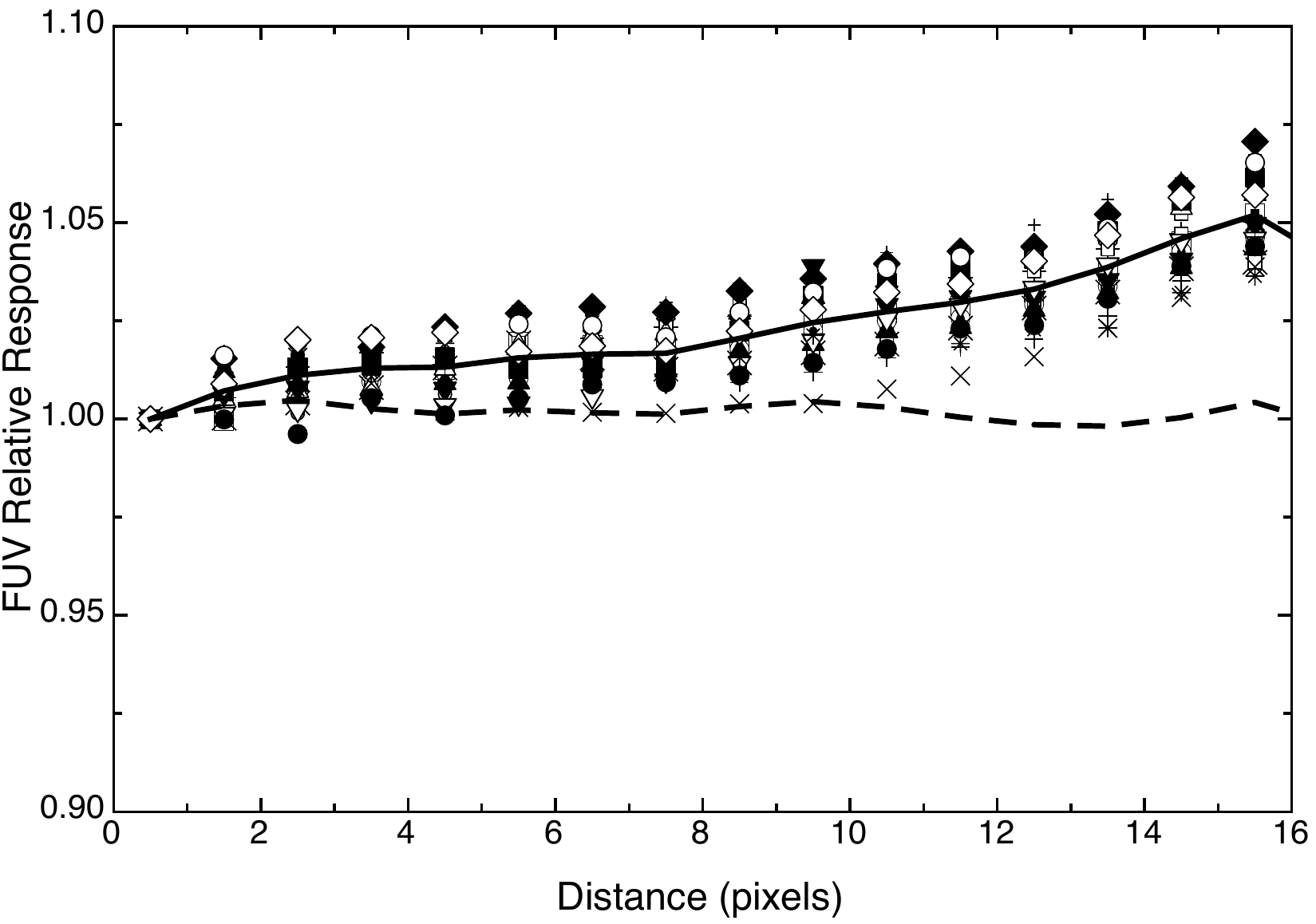}
\caption{Relative response over the FUV detector (a) with the annular response at different points in the Galaxy plotted in (b). The solid line in (b) represents the mean value of all observations and the dark dashed line represents the annular response after the uniformity correction.}
\label{fig:fuvnonuniformity}
\end{figure*}

\begin{figure*}[t]
\includegraphics[width=3in]{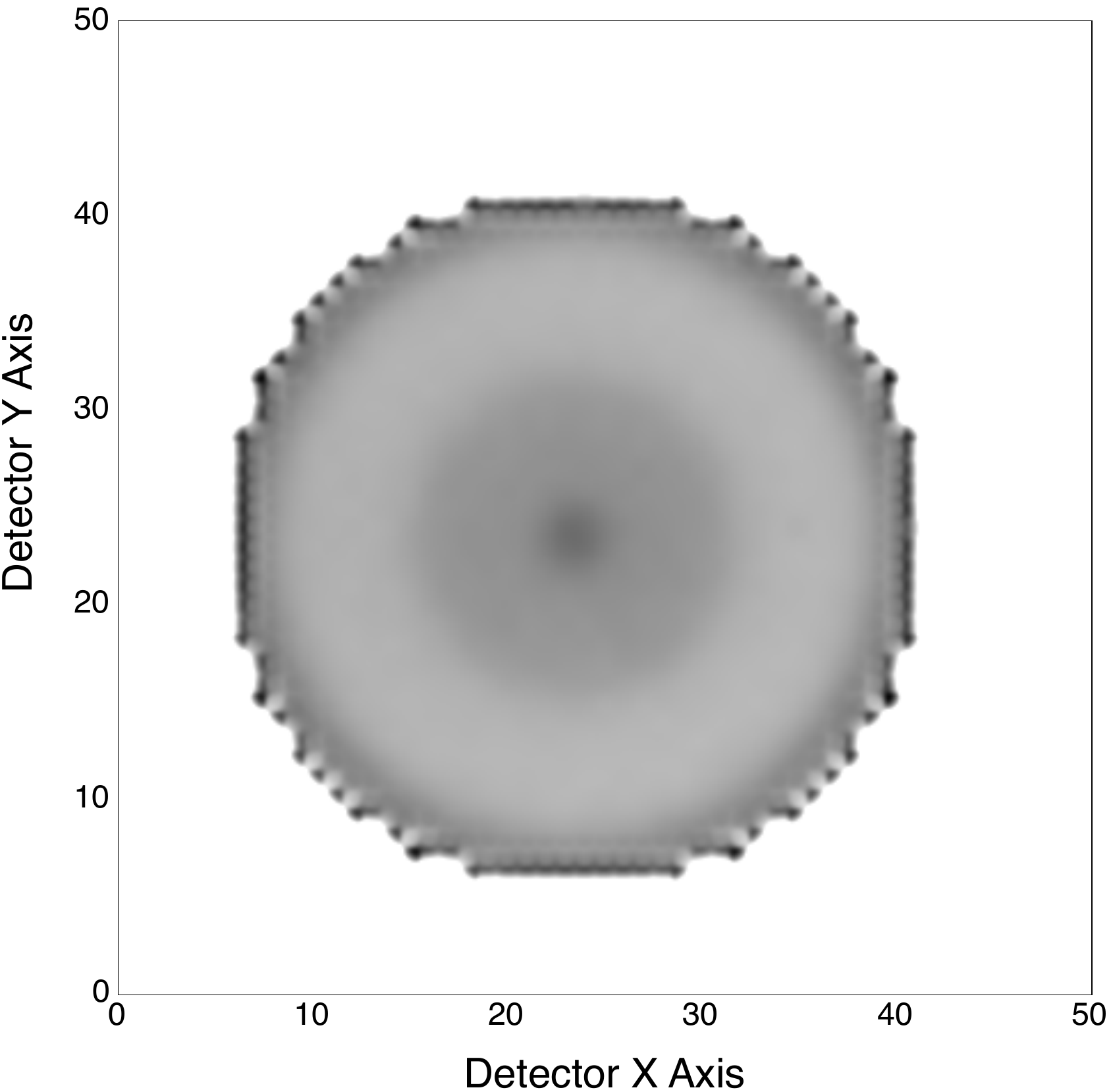}
\includegraphics[width=3in]{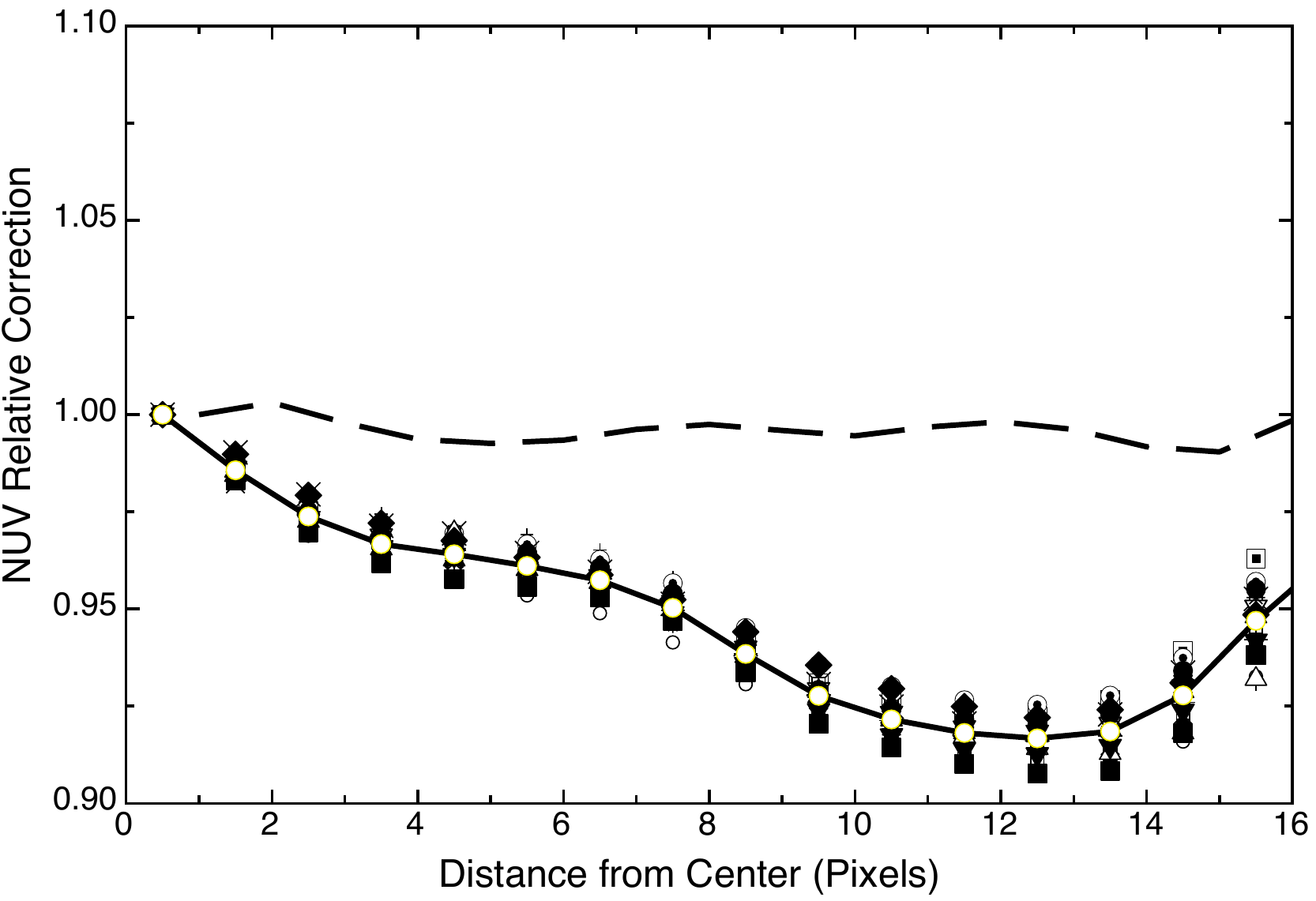}
\caption{Relative response over the NUV detector (a) with the annular response at different points in the Galaxy plotted in (b). The solid line in (b) represents the mean value of all observations and the dark dashed line represents the annular response after the uniformity correction.}
\label{fig:nuvnonuniformity}
\end{figure*}

\begin{table}[t]
\small
\caption{Non-uniformity Correction\label{table:unif}}
\begin{tabular}{llll}
\tableline
Exponent & FUV\tablenotemark{a} & NUV\tablenotemark{a}  \\
\tableline
0 & 0.9982 & 1.0211  \\
1 & $5.5540 \times 10^{-3}$ & $-4.3101 \times 10^{-2}$  \\
2 & $-7.2769 \times 10^{-5}$ & $1.3761 \times 10^{-2}$ \\
3 & $-1.4521 \times 10^{-4}$ & $-2.0514 \times 10^{-3}$   \\
4 & $1.8090 \times 10^{-5}$ &   $1.2921 \times 10^{-4}$ \\
5 & $-5.8485 \times 10^{-7}$ & $-2.8310 \times 10^{-6}$  \\
\tableline
\end{tabular}
\tablenotetext{a}{In both cases: $c = a0 + a1 * d + a2 * d^{2} + a3 *d^{3} + a4 * d^{4} + a5 * d^{5}$ where d is the distance from the center in binned pixels. $FUV = FUV/c_{FUV}$ and $NUV = NUV/c_{NUV}$. }
\end{table}

\citet{Hodges2014} speculated that there could be systematic effects in the uniformity correction for diffuse NUV sources because the white dwarfs used in the flat fielding have a much different spectral shape than the zodiacal light which forms a major part of the NUV background. I tested this for both bands by taking targets at different galactic latitudes and longitudes and adding the pixel counts together in detector coordinates (Fig. \ref{fig:fuvnonuniformity} and \ref{fig:nuvnonuniformity}). Although the diffuse light from the Galaxy certainly shows small scale variations, averaging over a large number of randomly selected observations should result in a uniform background over the \galex\ focal plane. The outer 20\% of the \galex\ images are affected by edge effects in the detectors and by scattered light from nearby stars and I have only used the central 0.5\degr\ radius (15 pixels radius) for my analysis. The FUV detector was relatively smooth with the edges being higher by about 5\% than the center (Fig. \ref{fig:fuvnonuniformity}) but the NUV detector showed a much stronger variation of up to 10\% (Fig. \ref{fig:nuvnonuniformity}). This shape of this nonuniformity is the same over the entire Galaxy from high latitude regions to low latitude regions, even though the relative contribution of the zodiacal light is much lower at low galactic latitudes indicating that the nonuniformity cannot be due to flat fielding issues. Morrissey (2014: personal communication) has suggested that the effect is rather due to scattering of the background from the shiny edge of the NUV detector. There is no such edge in the FUV detector. In either case, I have fit the profile with a $5^{th}$ order polynomial (Table \ref{table:unif}) which reduced the deviation to less than 2\% over the central 15 pixels (0.5\degr\ radius) of the field. Note that I have arbitrarily normalized the correction to the center of the image and that this correction is applied before the subtraction of the foreground.

\begin{figure*}[t]
\includegraphics[width=3in]{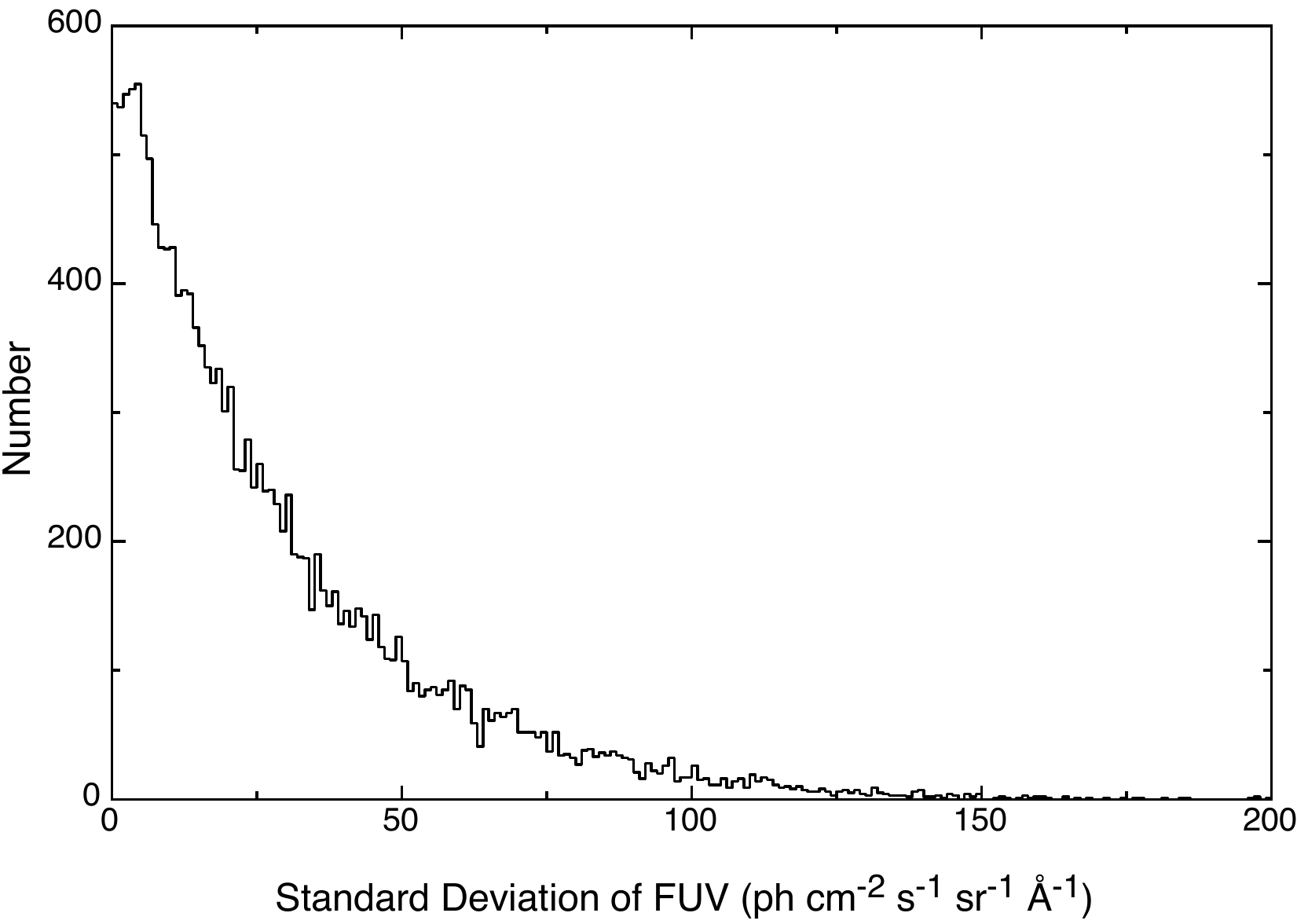}
\includegraphics[width=3in]{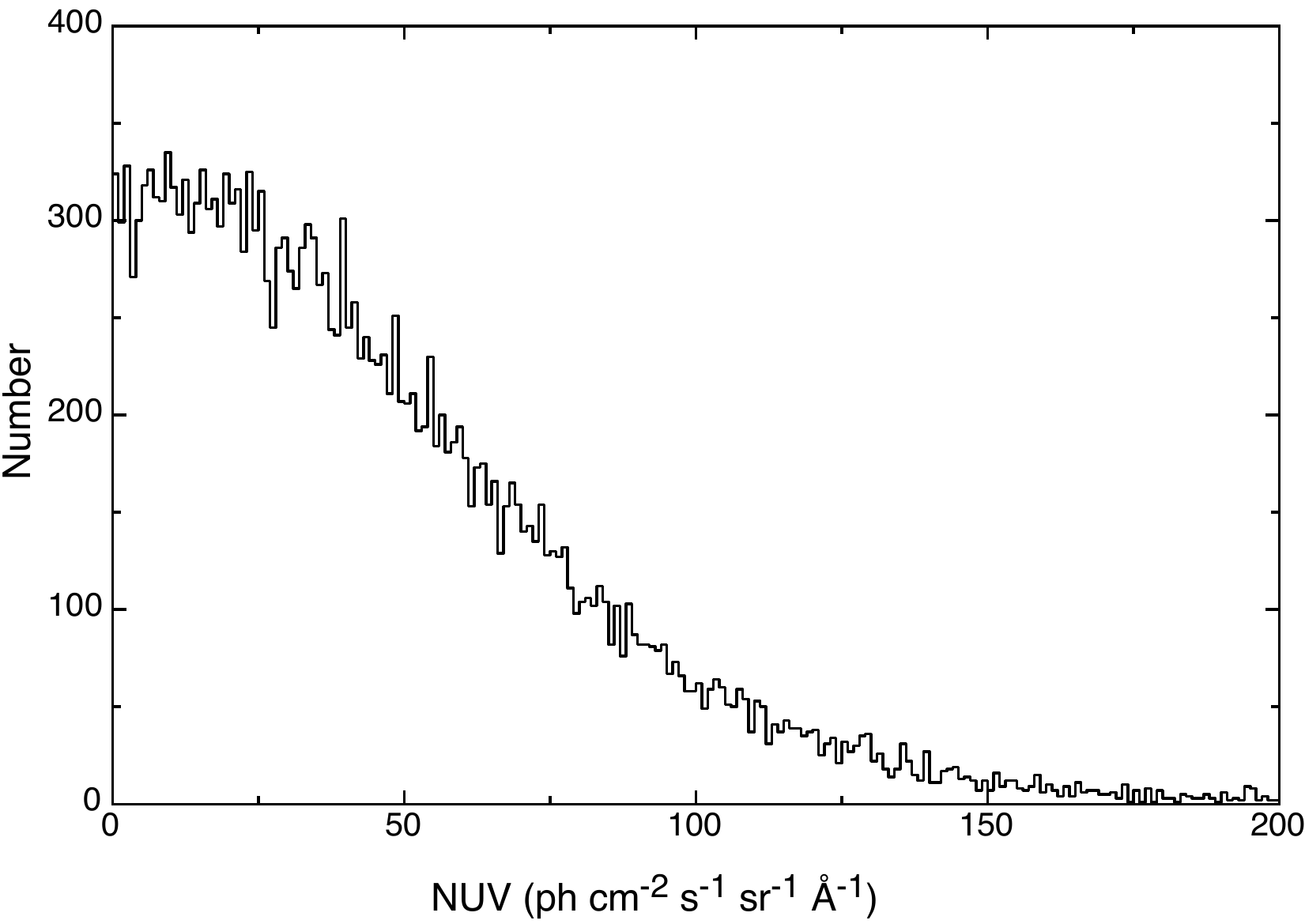}
\caption{Standard deviations in the derived FUV (a) and NUV (b) for each area of the sky with multiple visits.}
\label{fig:bkgdhistograms}
\end{figure*}

A significant part of the diffuse emission, particularly at high Galactic latitudes, is due to foreground sources --- airglow from geocoronal oxygen in both the FUV and NUV bands and zodiacal light in the NUV band only --- which are a function of the observing time and geometry. As discussed above, a single exposure (visit) is restricted to a maximum of about 1000 seconds because the spacecraft is in low Earth orbit and an observation may be made up of several visits separated by long periods of time. The majority of observations were made as part of the AIS survey and consist of a single visit of about 100 seconds in length. \citet{Murthy2014} derived empirical corrections to the airglow and the zodiacal light for each visit using the spacecraft housekeeping files (\textit{-scst}). These corrections will be invariant over the $1\degr$ usable \galex\ field of view and I subtracted them from each visit. In principle, the remaining diffuse emission should be entirely astrophysical and therefore should be consistent across visits for a given area in the sky. I have plotted the deviations over each visit that observed a single patch of sky in  Fig. \ref{fig:bkgdhistograms} finding a median deviation of about 20 \photu\ in the FUV and 40 \photu\ in the NUV. In most cases, these were different visits of a single observation but there were also multiple observations of a single location but, either way, they were a measure of the quality of the foreground correction. There were a few locations where there was a much larger deviation in the values with most of these being due to processing errors in the pipeline. These will be investigated individually in the future.

In partnership with the Mikulski Archive for Space Telescope (MAST), I have prepared and made available for public access several data files for the diffuse background at \url{http://archive.stsci.edu/prepds/uv-bkgd/}. The first of these is a set of files (Table \ref{table:filenames}) containing all the binned diffuse data (at 2\arcmin\ resolution) for each of the 100,864 visits. I have chosen to work at the visit level rather than the observation level because the foreground emission will vary between visits. For convenience, I have also provided another file with corrections at the observation level (described below). The files have been divided into latitude intervals of 15\degr\ with a cumulative total of 232,390,656 lines and a total data size of 35 GB. Each file is a space-delimited text file with the format listed in Table \ref{table:basedata}. Because the foreground emission may be different for each visit and so varies across an observation, I have used the individual visits for my input files. The first few columns (Columns 1 - 5) of the table are self-explanatory and are taken directly from the header of the original FITS file. The name of the original FITS can be easily reconstructed from the root in Column 1 by adding the appropriate extension: {\it fd-int.fits} or {\it nd-int.fits} for the image files and {\it xd-mcat.fits} for the merged catalog file. As discussed above, the diffuse file is made up of 2\arcmin\ pixels with each pixel comprising $80 \times\ 80$ pixels of the original image files and Columns 6 and 7 give the x and y position of these binned pixels. In practice, the \galex\ detectors suffer from edge effects and I have restricted my analysis to the central 0.5\degr\ (radius) of the image ($([x - 24]^{2} + [y - 24]^{2}) < 225$). Columns 8 and 9 are the Galactic coordinates for each pixel and 10 and 11 are the ecliptic coordinates of the pixel with Coumns 12 and 13 being the ecliptic coordinates of the Sun on the date of the visit. Columns 14 and 15 are the binned FUV and NUV values, respectively, corrected for the nonuniformity in the detector as discussed above. I have used the formula $2000 e^{-SA^2}$, where SA is the angle in radians between the Sun and the direction of the observation  \citep{Murthy2014} to calculate the airglow tabulated in Columns 16 and 17. I also showed in that paper that the zodiacal light in the UV (Column 18) follows its visible light distribution \citep{Leinert1998} but with a scale factor of 0.63 (modified slightly from the 0.65 cited by \citet{Murthy2014}). The final foreground-subtracted values of the diffuse FUV and NUV emission are listed in Columns 19 and 20 and form the main product from this work. The remaining columns are the median value for the background in the FUV and NUV bands (Columns 21 and 22) for that particular visit; the minimum value for the background across all visits (Columns 23 and 24); and the deviations in the median background across all visits (Columns 25 and 26). If there was only one visit covering a given field, I arbitrarily set the values in Columns 25 and 26 to 50 \photu\, . The last two columns (Column 27 and 28) contain the value of the 100 \micron\ emission and the predicted E(B-V) for the given coordinates from \citet{Schlegel1998}. In all columns, missing data were assigned a value of -9999. Programs to read these files are available from the MAST site.

\begin{figure*}[t]
\includegraphics[width=6in]{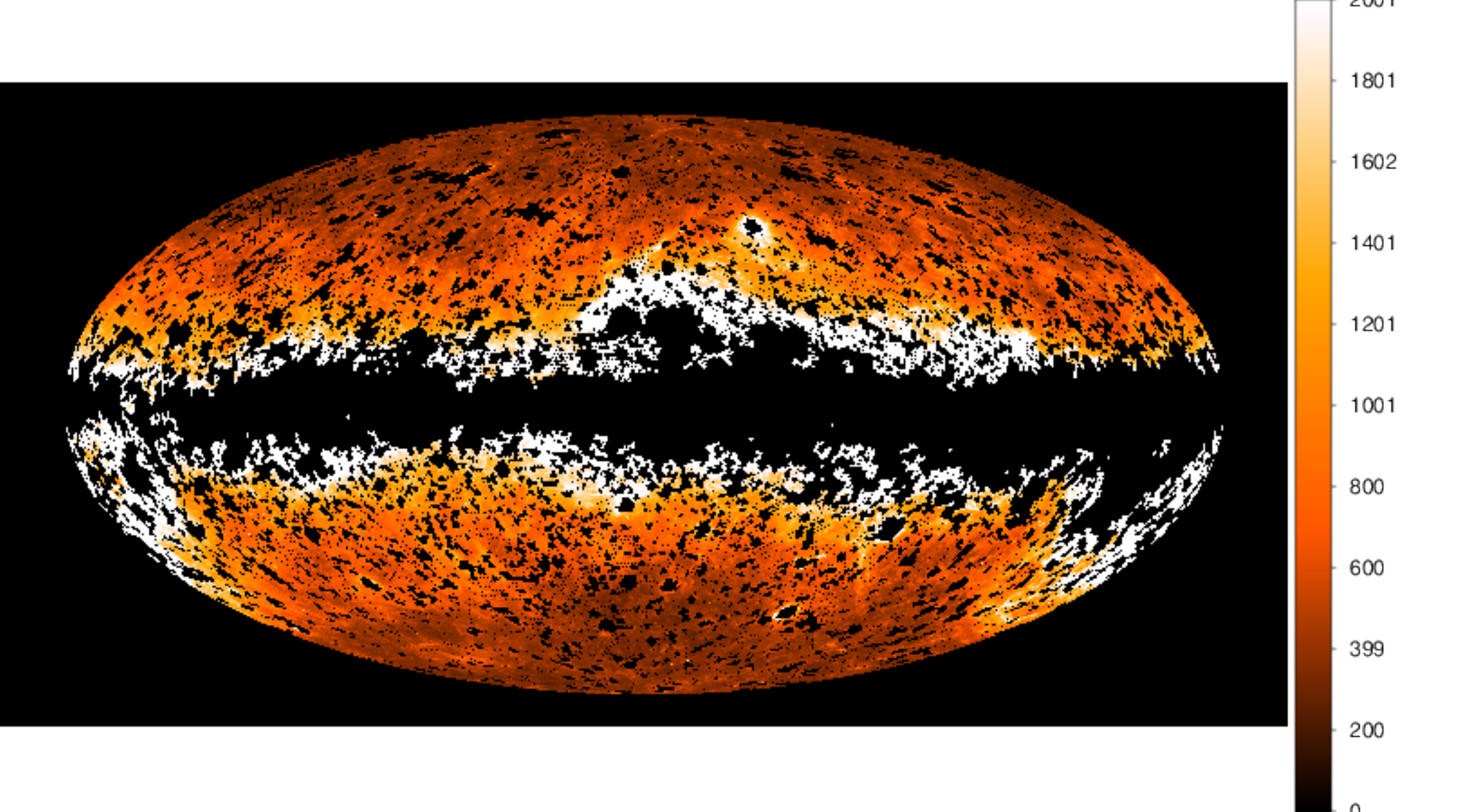}
\includegraphics[width=6in]{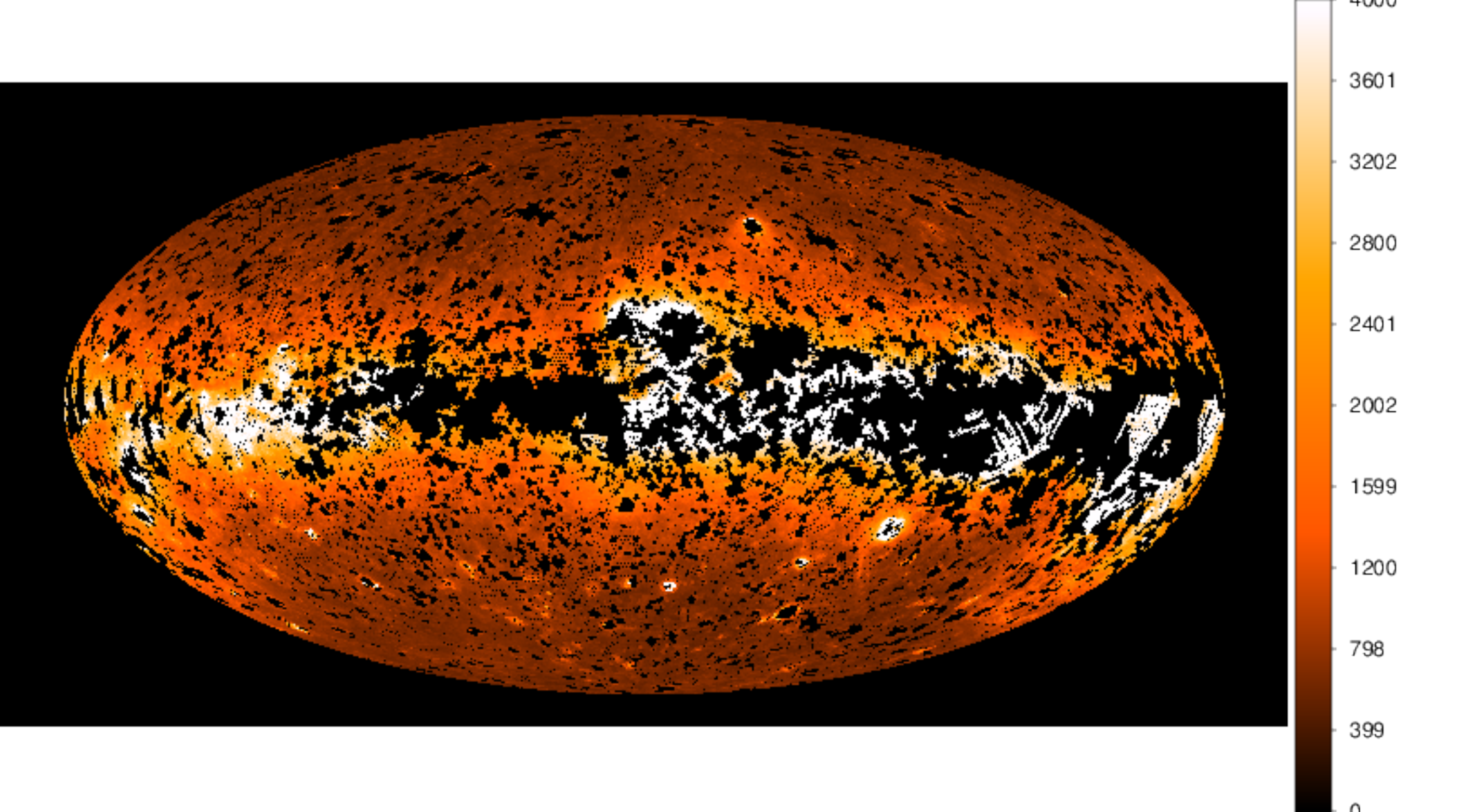}
\caption{The FUV (a) and NUV (b) backgrounds are shown. Note the different color scales as the NUV sky is much
brighter than the FUV sky. Color versions are available in the online journal.}
\label{fig:uvmaps}
\end{figure*}

The foreground subtracted values in Columns 19 and 20 have been derived using the same procedure that we have used in our past work but with a better accounting of the airglow and the zodiacal light. For ease of use, I have combined all these data to make Aitoff all-sky maps of the background in both the FUV and NUV bands (Fig. \ref{fig:uvmaps}). I have used a pixel size of 6\arcmin\ in these maps and have binned all values which fell within each pixel weighting each  by its exposure time. Because of edge effects in the \galex\ detectors, I have only used the central 0.5\degr\ (radius) of the field. The primary extension of each of the two files contains the diffuse background while the second extension contains the variance for each pixel (Columns 25 and 26 from Table \ref{table:basedata}). As discussed above, the standard deviation is typically less than 20 \photu\ in the FUV and 40 \photu\ in the NUV. Those areas with significantly larger deviations are generally due to problems in the pipeline extractions and will be investigated in detail in future work.

\begin{table}[h]
\small
\caption{Background Data File Names\label{table:filenames}}
\begin{tabular}{llll}
\tableline
GB\tablenotemark{a}  & Filename & No. of Rows & Size\tablenotemark{b}\\
\tableline
-90 -- -80 & \verb+hlsp_uv-bkgd_galex_diffuse_glat90-80S_fuv-nuv_v1_table.txt+ &  2196993& 0.3\\
-80 -- -70 & \verb+hlsp_uv-bkgd_galex_diffuse_glat80-70S_fuv-nuv_v1_table.txt+ &  6891683& 1.1\\
-70 -- -60 & \verb+hlsp_uv-bkgd_galex_diffuse_glat70-60S_fuv-nuv_v1_table.txt+ & 11117124& 1.7\\
-60 -- -50 & \verb+hlsp_uv-bkgd_galex_diffuse_glat60-50S_fuv-nuv_v1_table.txt+ & 17521838& 2.7\\
-50 -- -40 & \verb+hlsp_uv-bkgd_galex_diffuse_glat50-40S_fuv-nuv_v1_table.txt+ & 16406149& 2.6\\
-40 -- -30 & \verb+hlsp_uv-bkgd_galex_diffuse_glat40-30S_fuv-nuv_v1_table.txt+ & 26511792& 4.2\\
-30 -- -20 & \verb+hlsp_uv-bkgd_galex_diffuse_glat30-20S_fuv-nuv_v1_table.txt+ & 21512539& 3.5\\
-20 -- -10 & \verb+hlsp_uv-bkgd_galex_diffuse_glat20-10S_fuv-nuv_v1_table.txt+ & 11482259& 1.9\\
-10 --   0 & \verb+hlsp_uv-bkgd_galex_diffuse_glat10-00S_fuv-nuv_v1_table.txt+ &  6302181& 1.1\\
  0 --  10 & \verb+hlsp_uv-bkgd_galex_diffuse_glat00-10N_fuv-nuv_v1_table.txt+ &  6370046& 1.1\\
 10 --  20 & \verb+hlsp_uv-bkgd_galex_diffuse_glat10-20N_fuv-nuv_v1_table.txt+ & 13249318& 2.1\\
 20 --  30 & \verb+hlsp_uv-bkgd_galex_diffuse_glat20-30N_fuv-nuv_v1_table.txt+ & 16422219& 2.6\\
 30 --  40 & \verb+hlsp_uv-bkgd_galex_diffuse_glat30-40N_fuv-nuv_v1_table.txt+ & 18395057& 2.8\\
 40 --  50 & \verb+hlsp_uv-bkgd_galex_diffuse_glat40-50N_fuv-nuv_v1_table.txt+ & 18917725& 2.9\\
 50 --  60 & \verb+hlsp_uv-bkgd_galex_diffuse_glat50-60N_fuv-nuv_v1_table.txt+ & 15277170& 2.4\\
 60 --  70 & \verb+hlsp_uv-bkgd_galex_diffuse_glat60-70N_fuv-nuv_v1_table.txt+ & 10857896& 1.7\\
 70 --  80 & \verb+hlsp_uv-bkgd_galex_diffuse_glat70-80N_fuv-nuv_v1_table.txt+ & 10550629& 1.7\\
 80 --  90 & \verb+hlsp_uv-bkgd_galex_diffuse_glat80-90N_fuv-nuv_v1_table.txt+ &  2408038& 0.4\\
\tableline
\tablenotetext{a}{Range in Galactic latitude (in degrees) for file.}
\tablenotetext{b}{Approximate size in GB. May vary across machines.}
\end{tabular}
\end{table}

\begin{table}[t]
\small
\renewcommand{\arraystretch}{0.8}
\caption{Background Data File Format\label{table:basedata}}
\begin{tabular}{llll}
\tableline
Column No. & Name & Format & Description\\
\tableline
1 & Name & String & Name of the \galex\ pipeline file\\
2 & Date & String & Observation date from \galex\ pipeline file\\
3 & Time & String & Observation time from \galex\ pipeline file\\
4 & FUV Exp. Time & Integer & Total exposure time in the FUV band (s)\\
5 & NUV Exp. Time & Integer & Total exposure time in the NUV band (s)\\
6 & X &  Integer & Binned pixel in X\tablenotemark{a}\\
7 & Y &  Integer & Binned pixel in Y\tablenotemark{a}\\
8 & GL &  Float & Galactic Longitude\tablenotemark{b}\\
9 & GB &  Float & Galactic Latitude\tablenotemark{b}\\
10 & ECL & Float & Ecliptic Longitude\tablenotemark{b}\\
11 & ECB & Float & Ecliptic Latitude\tablenotemark{b}\\
12 & ECL$_{Sun}$ & Float & Ecliptic Longitude of Sun\tablenotemark{b}\\
13 & ECB$_{Sun}$ & Float & Ecliptic Latitude of Sun\tablenotemark{b}\\
14 & FUV$_{orig}$ & Integer & FUV flux from \galex \tablenotemark{c}\\
15 & NUV$_{orig}$ & Integer & NUV flux from \galex \tablenotemark{c}\\
16 & FUV$_{AG}$ & Integer & FUV airglow contribution\tablenotemark{c,d}\\
17 & NUV$_{AG}$ & Integer & NUV airglow contribution\tablenotemark{c,d}\\
18 & NUV$_{ZL}$ & Integer & NUV zodiacal light contribution\tablenotemark{c,d}\\
19 & FUV$_{Final}$ & Integer & Final FUV background\tablenotemark{c}\\
20 & NUV$_{Final}$ & Integer & Final NUV background\tablenotemark{c}\\
21 & FUV$_{Med}$ & Integer & Median FUV background for the observation\tablenotemark{c}\\
22 & NUV$_{Med}$ & Integer & Median NUV background for the observation\tablenotemark{c}\\
23 & FUV$_{Min}$ & Integer & Minimum value for the FUV background across observations\tablenotemark{c}\\
24 & NUV$_{Min}$ & Integer & Minimum value for the NUV background across observations\tablenotemark{c}\\
25 & $\delta$FUV & Integer & Standard deviation in FUV across observations\tablenotemark{c}\\
26 & $\delta$NUV & Integer & Standard deviation in NUV across observations\tablenotemark{c}\\
27 & 100 \micron\ & Float & 100 \micron\  emission (MJy sr$^{-1}$)\tablenotemark{e}\\
28 & E(B-V) & Float & E(B-V) (magnitudes)\tablenotemark{e}\\

\tableline
\end{tabular}
\tablenotetext{a}{One binned pixel in the diffuse file is 80 pixels $\times$ 80 pixels in the original data file.}
\tablenotetext{b}{Degrees}
\tablenotetext{c}{\photu\ May be Long in some implementations. }
\tablenotetext{d}{\citet{Murthy2014}}
\tablenotetext{e}{\citet{Schlegel1998}}
\end{table}
\clearpage 

\section{Galactic Trends}

\begin{figure*}[t]
\includegraphics[width=3in]{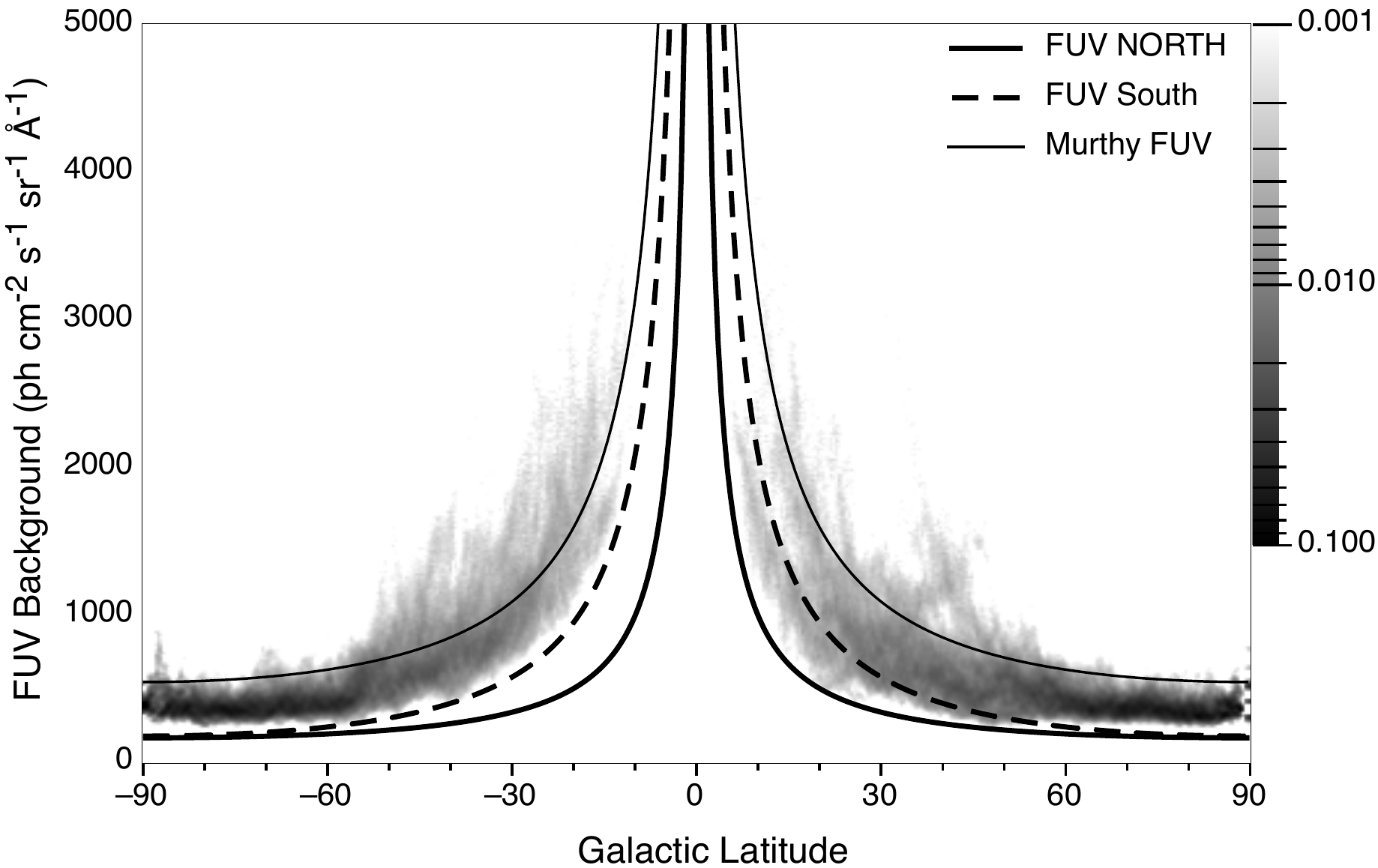}
\includegraphics[width=3in]{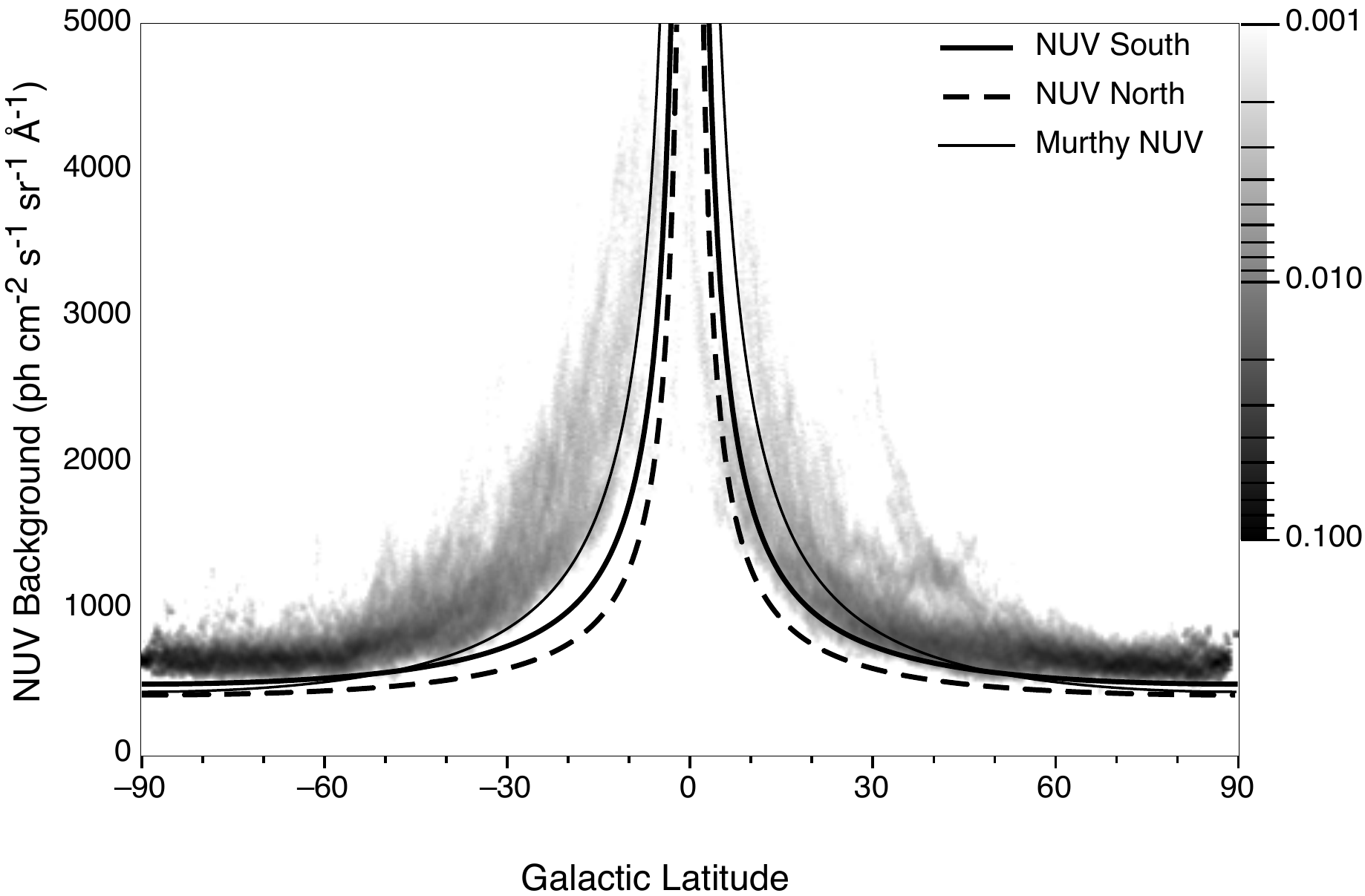}
\caption{Density plots of FUV (a) and NUV (b) backgrounds are shown. The shading is normalized to the number of observations in each latitude interval. Best fit cosecant laws are shown for each hemisphere (see text). The equivalent cosecant law from \citet{Murthy2010} is shown for comparison. }
\label{fig:cscplots}
\end{figure*}

\begin{table}[t]
\small
\caption{Cosecant Law for Diffuse Background\label{table:csclaws}}
\begin{tabular}{llll}
\tableline
 Region& Constant\tablenotemark{a} & Slope\tablenotemark{a} & r\tablenotemark{b}  \\
\tableline
FUV South ($-50 < b < -15$) & -205.5 & 401.8 & 0.985\\
NUV South ($-50 < b < -15$) & 66.7 & 356.3 & 0.971\\
FUV North ($15 < b < 50$) & 93.4 & 133.2 & 0.966\\
NUV North ($15 < b < 50$) & 257.5 & 185.1 & 0.971\\

\tableline
\end{tabular}
\tablenotetext{a}{\photu}
\tablenotetext{b}{Correlation coefficient between background and cosecant of Galactic latitude.}

\end{table}

\begin{figure*}[ht]
\begin{minipage}[t]{0.45\linewidth}
\includegraphics[width=3in]{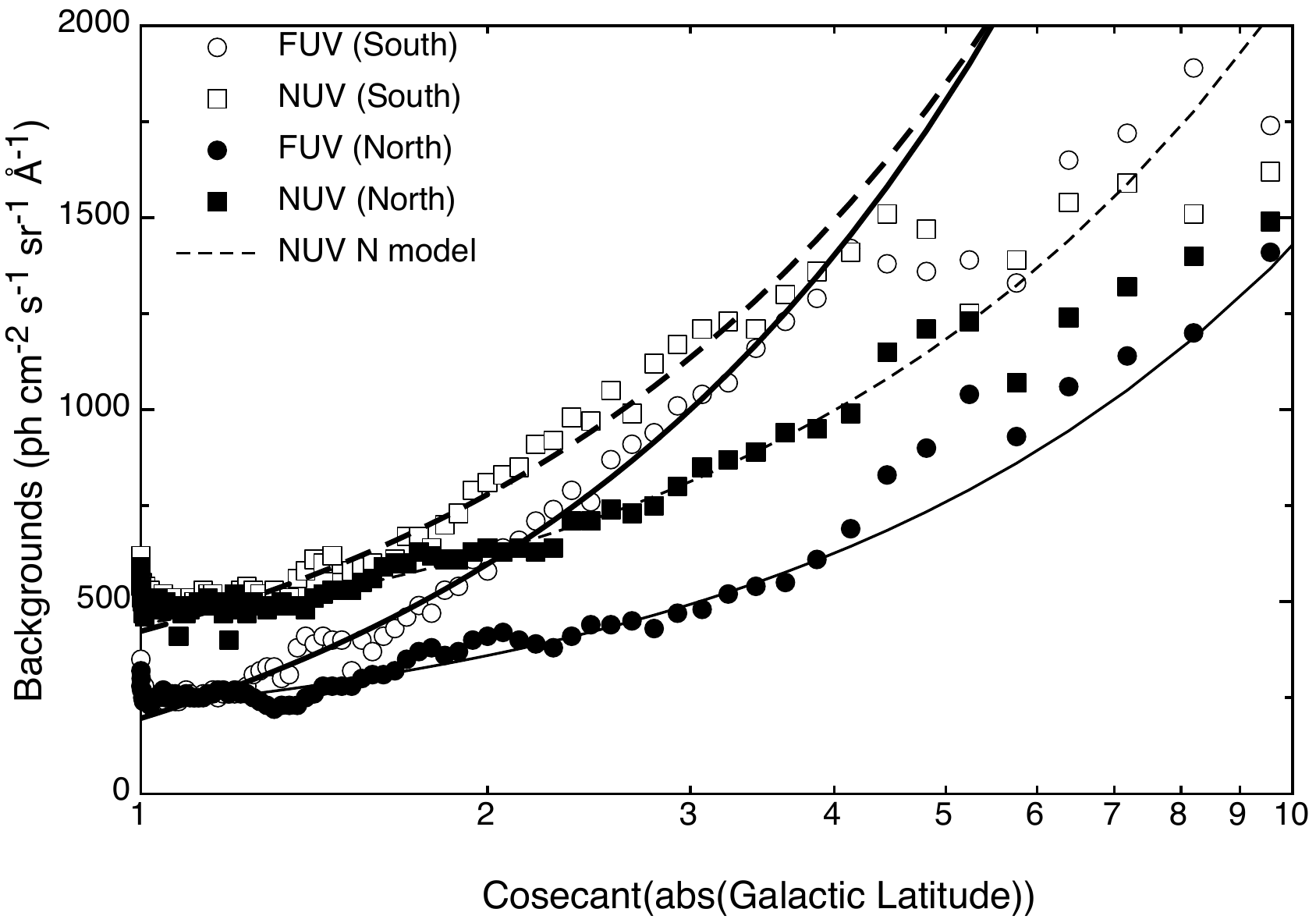}
\caption{The minimum FUV (squares) and NUV (circles) backgrounds are shown in the Southern (solid symbols)
and Northern (unfilled symbols) hemispheres with the best linear fits between latitudes of 15\degr\ and 50\degr \
in each hemisphere.}
\label{fig:csclaw}
\end{minipage}
\quad
\begin{minipage}[t]{0.45\linewidth}
\includegraphics[width=3in]{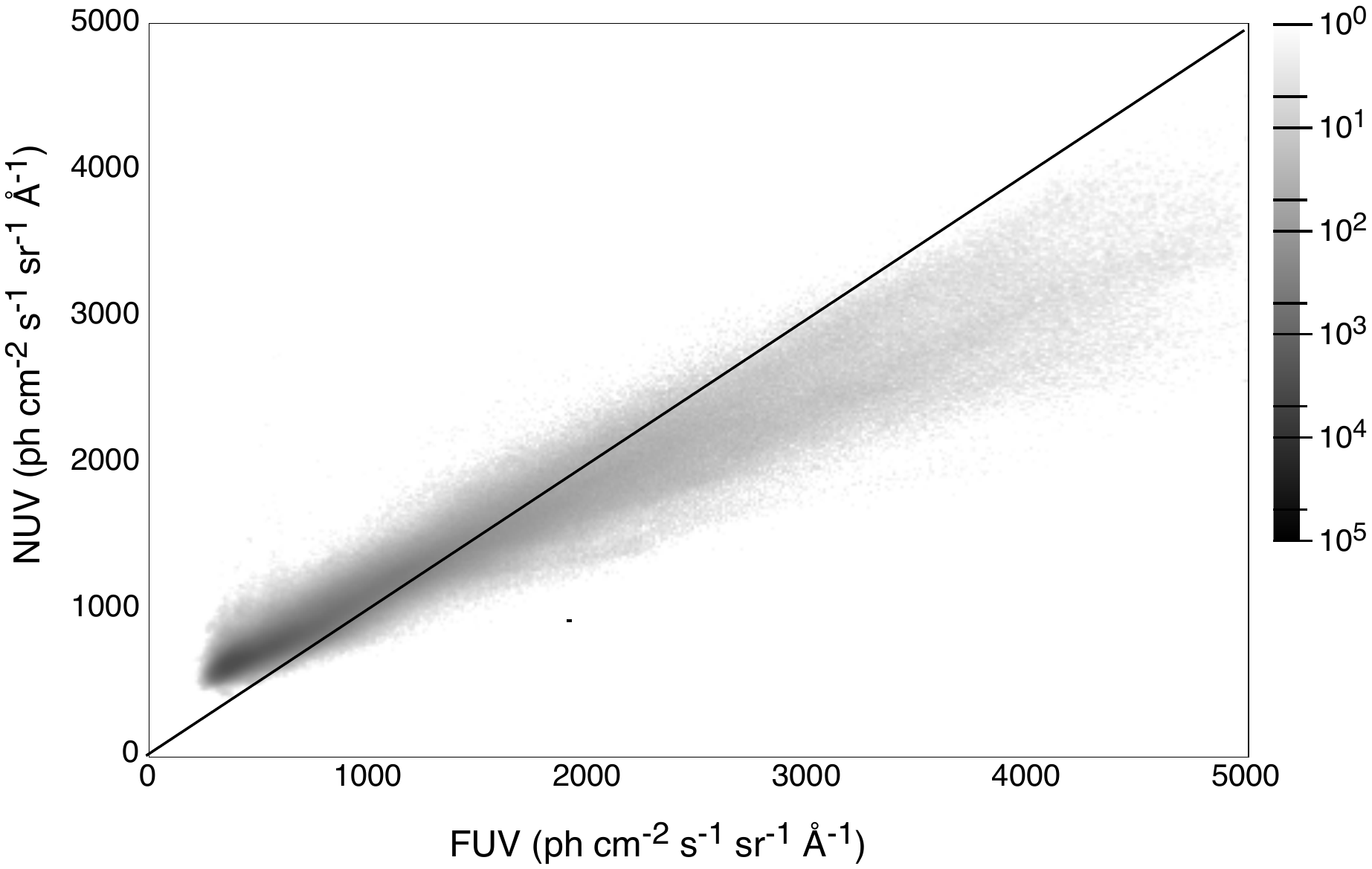}
\caption{Density plot of NUV versus FUV.}
\label{fig:fuvnuv}
\end{minipage}
\end{figure*}

The trends in this work are essentially the same as those in \citet{Murthy2010} and \citet{Hamden2013} as would be expected given that the source of the data is the \galex\ spacecraft in all cases, albeit processed differently and at different resolutions. \citet{Murthy2010} found that both the FUV and NUV followed cosecant laws with respective slopes of 545 and 433 \photu, close to the 540 \photu\ found by \citet{Wright1992} from DE-1 ({\it Dynamics Explorer}) data and the 412 \photu\ found by \citet{Seon2011}. I have plotted the density plots for the FUV and NUV background in Fig. \ref{fig:cscplots} and the inner envelope of the UV fluxes in Fig. \ref{fig:csclaw}. Although there is an excellent linear correlation between the UV fluxes and the cosecant of the Galactic latitude (Table \ref{table:csclaws}) at mid-latitudes, the diffuse radiation is asymmetric about the Galactic plane with different slopes to the fit in different hemispheres. This asymmetry is readily apparent in Fig. \ref{fig:cscplots} and, with hindsight, in Fig. 2 of \citet{Murthy2010}. The UV emission is a complex convolution of the stellar radiation field and the dust distribution, possibly with other contributors, and these trends will be investigated in further work.

\begin{figure*}[t]
\includegraphics[width=3in]{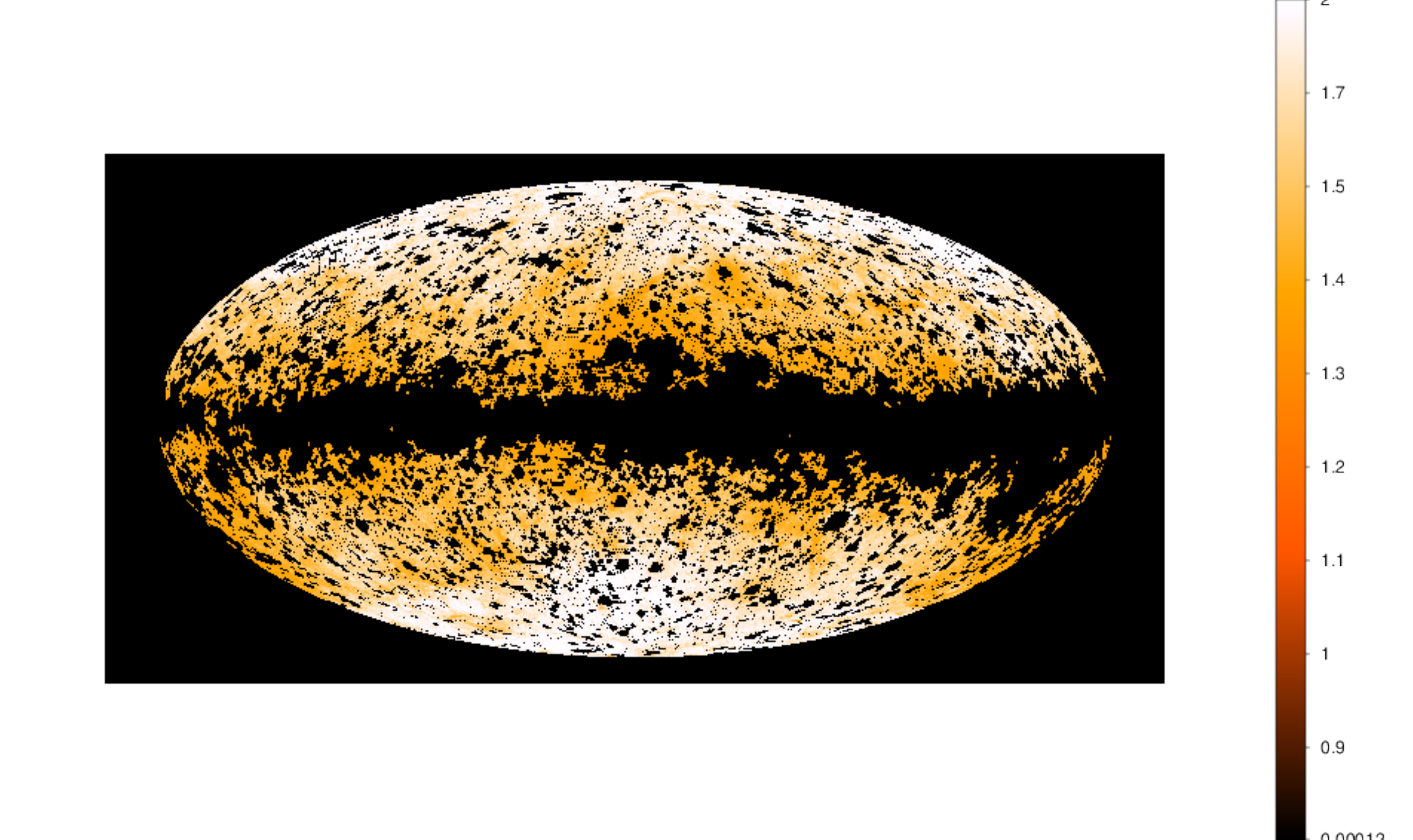}
\includegraphics[width=3in]{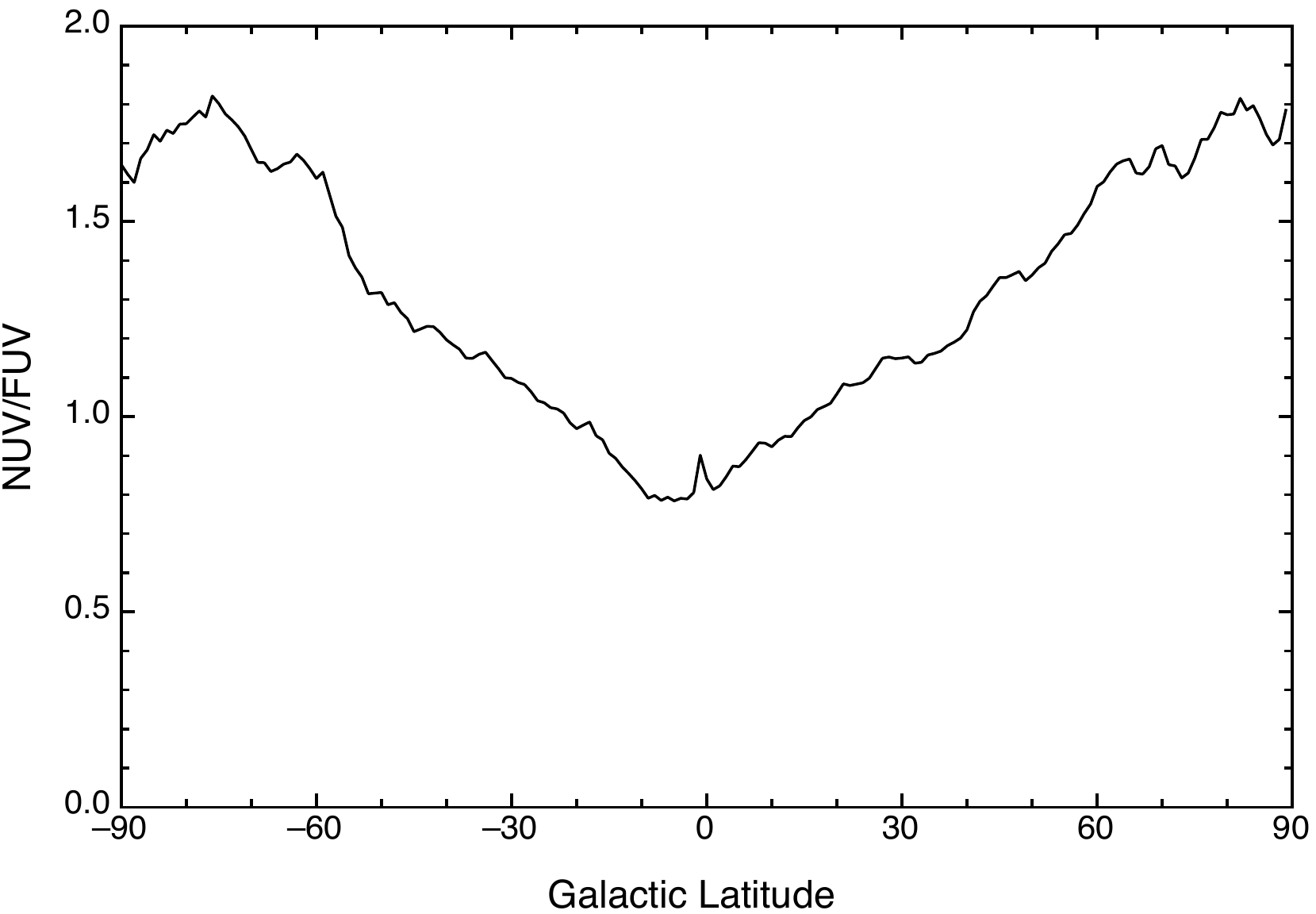}
\caption{Aitoff map of NUV/FUV ratio (a) with latitudinal dependence of the median ratio in (b).}
\label{fig:fuvnuvratio}
\end{figure*}

\begin{figure*}[t]
\includegraphics[width=3in]{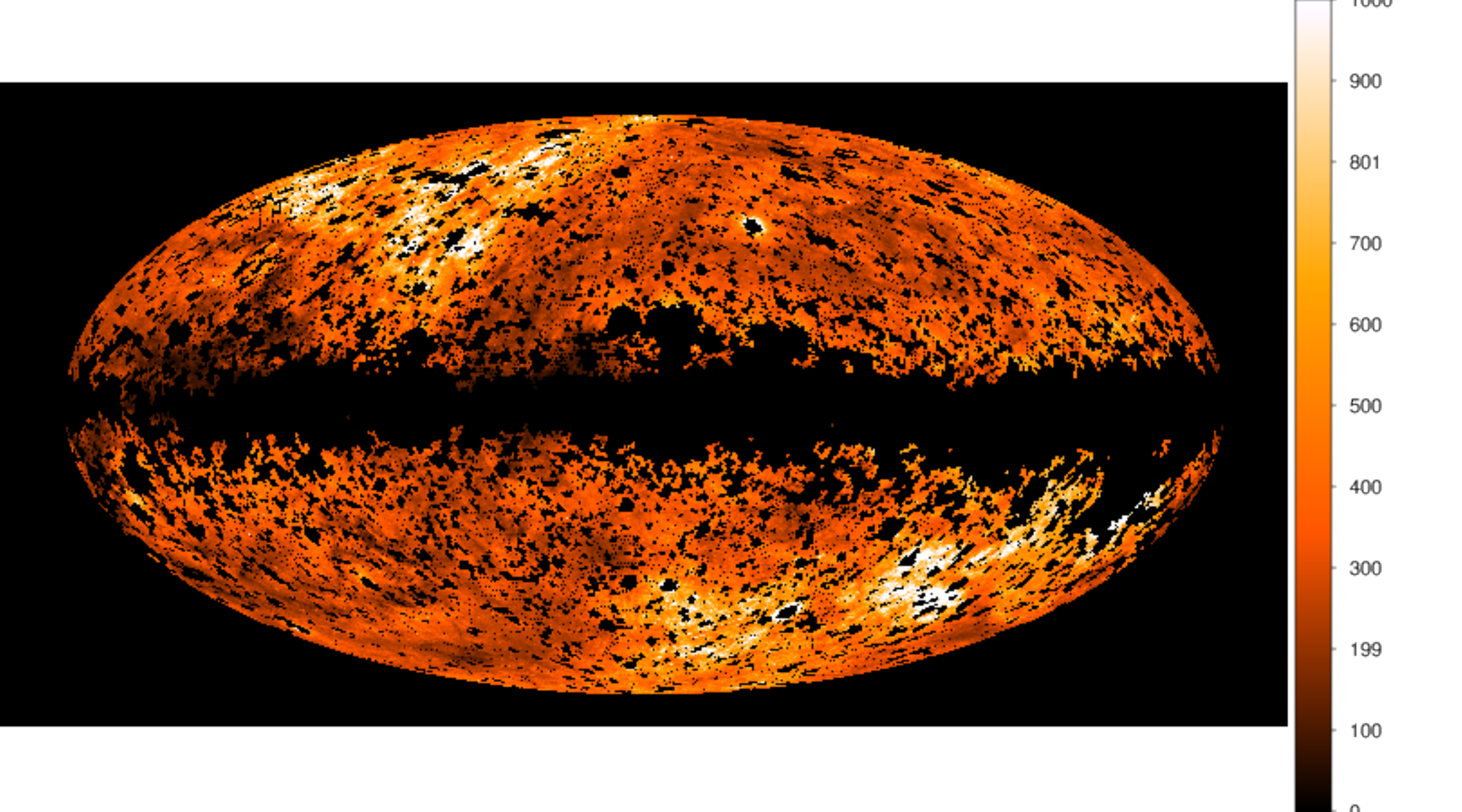}
\includegraphics[width=3in]{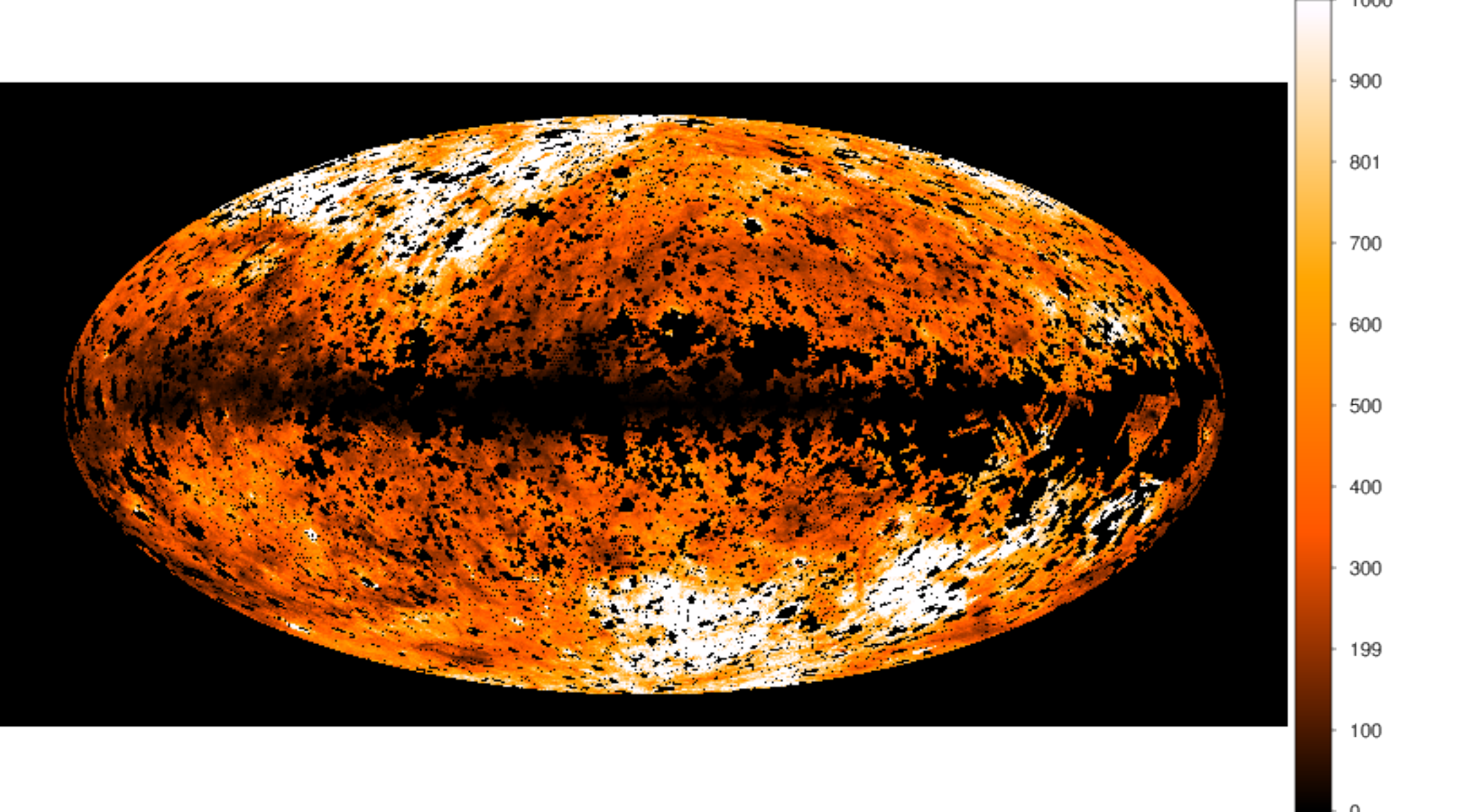}
\caption{Aitoff map of FUV/100 \micron\ and NUV/100 \micron\ ratios.}
\label{fig:uvirratio}
\end{figure*}

\begin{figure*}[h]
\begin{minipage}[t]{0.45\linewidth}
\includegraphics[width=2.8in]{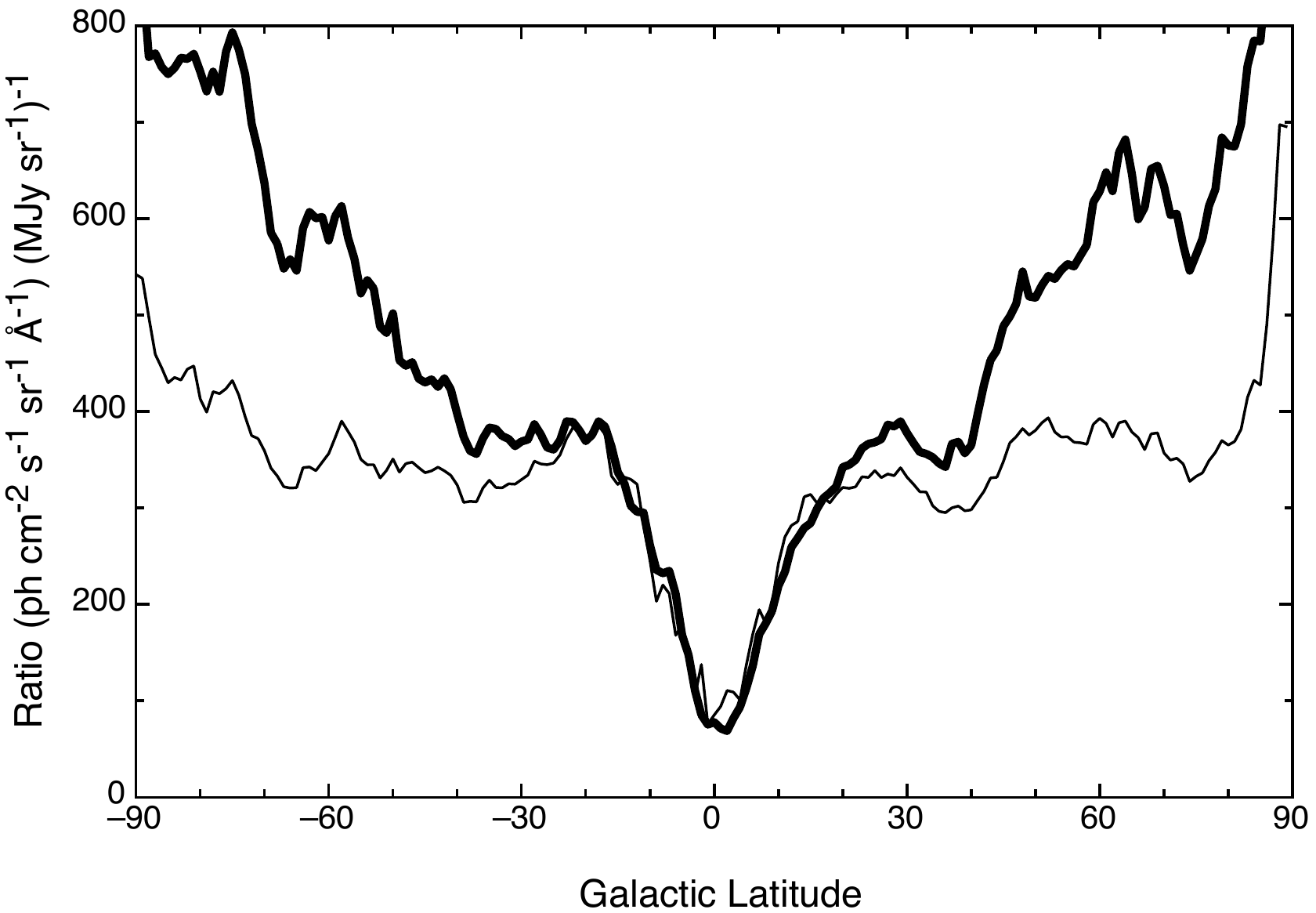}
\caption{Median ratio of FUV (thin line) and NUV (thick line) with the 100 \micron\ flux in \photu\ (MJy sr$^{-1}$)$^{-1}$ as a function of latitude.}
\label{fig:uvirratlat}
\end{minipage}
\quad
\begin{minipage}[t]{0.45\linewidth}
\includegraphics[width=2.8in]{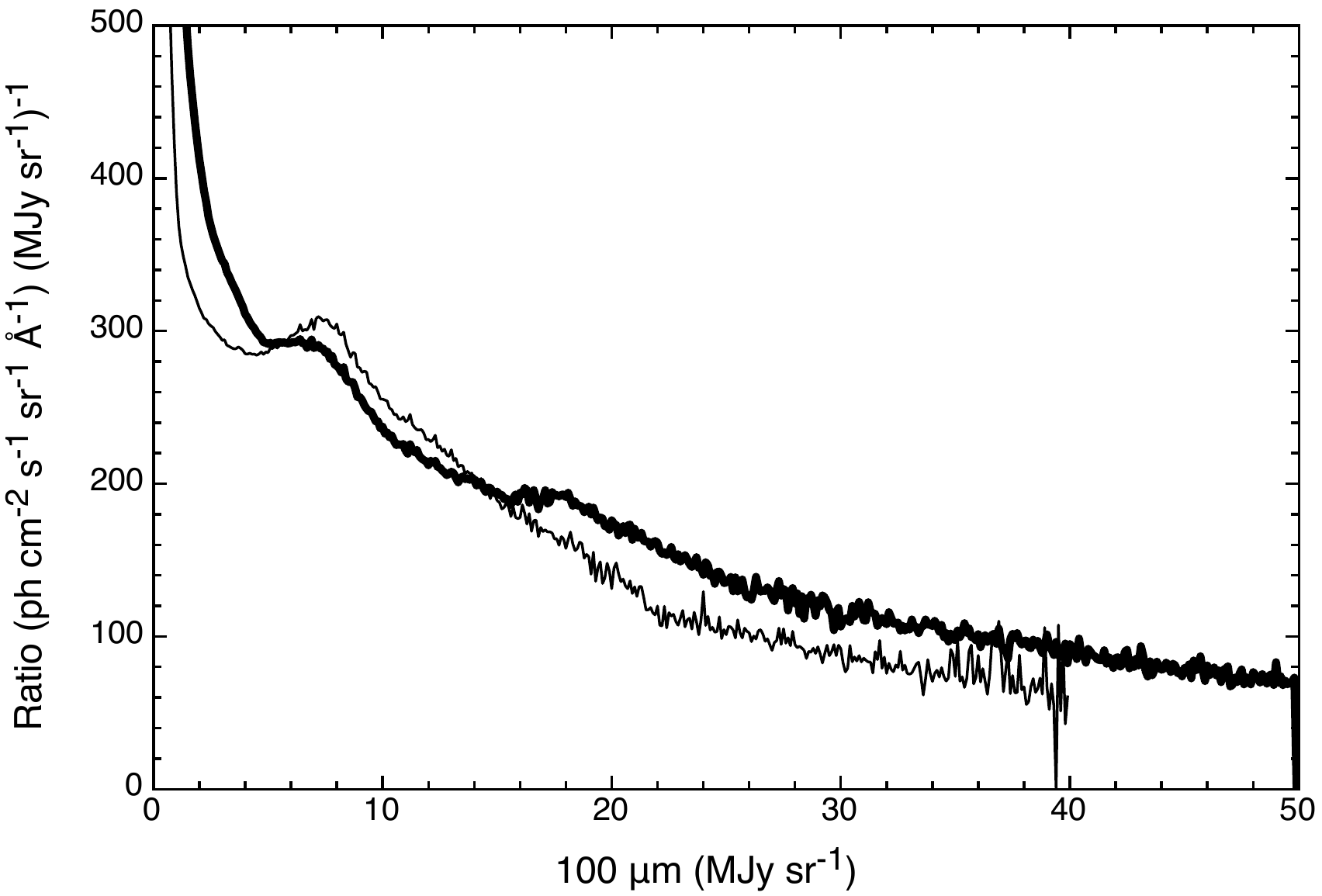}
\caption{Median ratio of FUV (thin line) and NUV (thick line) with the 100 \micron\ flux in \photu\ (MJy sr$^{-1}$)$^{-1}$ as a function of the 100 \micron\ flux.}
\label{fig:uvirratir}
\end{minipage}
\end{figure*}

The FUV and NUV emission are well-correlated (Fig. \ref{fig:fuvnuv}) with a linear correlation coefficient of 0.94. However, this is misleading as there is a clear latitudinal dependence of the FUV/NUV ratio with the ratio being highest at the poles (Fig. \ref{fig:fuvnuvratio}). A natural explanation for this might be that fluorescence in the Werner band of molecular hydrogen \citep{Sternberg1989} contributes to the FUV at low latitudes but not at high latitudes. However, there are other possibilities including a differing population of stars contributing in each band combined with different optical properties of the grains in the two bands. A comprehensive multi-wavelength modeling is required to separate the different contributors.

The FUV/100 \micron\ and NUV/100 \micron\ ratios show considerable variation over the sky (Fig. \ref{fig:uvirratio}) ranging from between 200 and 800 \photu\ (MJy sr$^{-1}$)$^{-1}$. The ratio drops at low latitudes (Fig. \ref{fig:uvirratlat}) but this is likely due to the increase in the 100 \micron\ emission in the Galactic disk. The optical depth is much higher in the UV than in the IR and hence the thermal emission from dust in the IR will continue to rise after the UV radiation saturates. This is seen in Fig. \ref{fig:uvirratir} where I have plotted the UV/IR ratio as a function of the 100\micron\ emission with the ratio dropping significantly as the IR emission increases.
\clearpage

\section{Additional Data Files}

\begin{table}[t]
\small
\caption{\galex\ Position File Format\label{table:sclong}}
\begin{tabular}{llll}
\tableline
Column No. & Name & Format & Description  \\
\tableline
1 & Year & Integer & Year of Observation\tablenotemark{a}  \\
2 & Month & Integer & Month of Observation\tablenotemark{a} \\
3 & Day & Integer & Day of Observation\tablenotemark{a} \\
4 & Hour & Float & Hour of Observation\tablenotemark{a} \\
5 & S/C Longitude  & Float & Longitude of spacecraft on the Earth\tablenotemark{b} \\
6 & S/C Longitude Rate & Float & Rate of change in longitude\tablenotemark{c} \\
7 & S/C Latitude &  Float & Latitude of spacecraft on the Earth\tablenotemark{b} \\
8 & S/C Latitude Rate &  Float & Rate of change in latitude\tablenotemark{c} \\

\tableline
\end{tabular}
\tablenotetext{a}{UT.}
\tablenotetext{b}{Degrees.}
\tablenotetext{c}{Degrees s$^{-1}$ }
\end{table}

I have made two further data files available. The first was created by \citet{Murthy2014} as part of an effort to characterize the foreground emission which was partly due to airglow. I used the spacecraft housekeeping files in that work which included the TEC (Total Event Count) as a function of observation time. Because the airglow turned out to be related to the time from local midnight, I had to convert the observation time from UT (as given in the archived data) to local spacecraft time. Unfortunately, there was no readily apparent way to do this from the information provided in the archive files. I therefore obtained the spacecraft TLEs (Two Line Elements) from Space-Track.org (\url{https://www.space-track.org}) from which I calculated the longitude and latitude of the \galex\ spacecraft using STK (\url{http://www.agi.com/default.aspx}). These are tabulated in a file (\verb+hlsp_uv-bkgd_galex_telemetry_telescope_fuv-nuv_v1_table.txt+) with 5,580,101 rows and with the columns listed in Table \ref{table:sclong}.

\begin{table}[t]
\small
\caption{\galex\ Observation Level Corrections\label{table:obsback}}
\begin{tabular}{llll}
\tableline
Column No. & Name & Format & Description  \\
\tableline
1 & File Name & String & Root for \galex\ observation.  \\
2 & Visits & Integer & Number of visits in the observation. \\
3 & FUV Back & Float & Median value of foreground subtracted diffuse flux\tablenotemark{a} \\
4 & NUV Back & Float & Median value of foreground subtracted diffuse flux\tablenotemark{a} \\
5 &  FUV AG & Float & Predicted airglow emission\tablenotemark{a} \\
6 & NUV AG & Float & Predicted airglow emission\tablenotemark{a} \\
7 & NUV ZL &  Float & Predicted zodiacal light contribution\tablenotemark{a} \\

\tableline
\end{tabular}
\tablenotetext{a}{\photu.}
\end{table}

The final data file (\verb+hlsp_uv-bkgd_galex_foregrounds_allsky_fuv-nuv_v1_table.txt+)is a set of foreground corrections for each of the observation level files, where a single  observation may be made up of multiple visits. The airglow and zodiacal light may be different for each visit and will be averaged for the total observation. I have created a look-up table where the foregrounds and the diffuse cosmic background are tabulated for each observation with a format documented in Table \ref{table:obsback}. Column 1 is the root for the observation and Column 2 is the number of visits in each observation. Columns 3 and 4 are the median values of the foreground-subtracted diffuse emission; ie., they represent the astrophysical background in the field. The remaining three columns include, in order, the effective airglow emission in the FUV and NUV bands and the predicted zodiacal light in the NUV band, calculated by taking the fluxes from each of the visit files and adding them together weighted by the exposure time.

\section{Conclusion}

I have processed the entire GR6/GR7 release of \galex\ to extract the diffuse cosmic background in each field. These data have been made available as part of the MAST HLSP archive (\url{http://archive.stsci.edu/prepds/uv-bkgd/}) as a set of ASCII files along with programs to read and bin the data as desired. I have also produced and made available Aitoff maps of the sky with 6\arcmin\ resolution in each of the FUV and NUV bands. I am now using these data for my own studies of the diffuse UV radiation and will continue to update the files as needed both in the MAST archive and on my own website at \url{http://www.iiap.res.in/personnel/murthy/Jayant\_Murthy/Home.html}.

\acknowledgments

This work would not have been possible without the many insights into the \galex\ instrument and data provided by Patrick Morrissey, help with accessing the archived data by Bernie Shiao, and help in getting the data into the right format by Scott Fleming. An anonymous referee clarified many of the concepts in this work. This research has made use of NASA's Astrophysics Data System Bibliographic Services. 

Some of the data presented in this paper were obtained from the Mikulski Archive for Space Telescopes (MAST). STScI is operated by the Association of Universities for Research in Astronomy, Inc., under NASA contract NAS5-26555. Support for MAST for non-HST data is provided by the NASA Office of Space Science via grant NNX13AC07G and by other grants and contracts.


\begin{thebibliography}{}

\bibitem[Bianchi(2014)]{Bianchi2014} Bianchi, L.\ 2014, 
arXiv:1404.4882 

\bibitem[Bianchi et al.(2014)]{BCS2014} Bianchi, L., Conti, A., 
\& Shiao, B.\ 2014, Advances in Space Research, 53, 900 

\bibitem[Bowyer(1991)]{Bowyer1991} Bowyer, S.\ 1991, \araa, 29, 59 

\bibitem[Choi et al.(2013)]{Choi2013} Choi, Y.-J., Min, K.-W., 
Seon, K.-I., et al.\ 2013, \apj, 774, 34 

\bibitem[Edelstein et al.(2006)]{Edelstein2006} Edelstein, J., Min, 
K.-W., Han, W., et al.\ 2006, \apjl, 644, L153 

\bibitem[Finkbeiner(2003)]{Finkbeiner2003} Finkbeiner, D.~P.\ 2003, 
\apjs, 146, 407

\bibitem[Hamden et al.(2013)]{Hamden2013} Hamden, E.~T., 
Schiminovich, D., \& Seibert, M.\ 2013, \apj, 779, 180 

\bibitem[Hayakawa et 
al.(1969)]{Hayakawa1969} Hayakawa, S., Yamashita, K., \& Yoshioka, S.\ 1969, \apss, 5, 493

\bibitem[Henry(1991)]{Henry1991} Henry, R.~C.\ 1991, \araa, 29, 89 

\bibitem[Hodges-Kluck 
\& Bregman(2014)]{Hodges2014} Hodges-Kluck, E., \& Bregman, J.\ 2014, arXiv:1401.4170 

\bibitem[Kalberla et 
al.(2005)]{Kalberla2005} Kalberla, P.~M.~W., Burton, W.~B., Hartmann, D., et al.\ 2005, \aap, 440, 775

\bibitem[Leinert et 
al.(1998)]{Leinert1998} Leinert, C., Bowyer, S., Haikala, L.~K., et al.\ 1998, \aaps, 127, 1 

\bibitem[Martin et al.(2005)]{Martin2005} Martin, D.~C., Fanson, 
J., Schiminovich, D., et al.\ 2005, \apjl, 619, L1 

\bibitem[Morrissey et al.(2007)]{Morrissey2007} Morrissey, P., Conrow, T., Barlow, T. A., et al. 2007 \apjs 173, 682

\bibitem[Murthy(2009)]{Murthy2009} Murthy, J.\ 2009, \apss, 320, 21

\bibitem[Murthy(2014)]{Murthy2014} Murthy, J.\ 2014, \apss, 349, 165 

\bibitem[Murthy 
\& Henry(2011)]{Murthy2011} Murthy, J., \& Henry, R.~C.\ 2011, \apj, 734, 13 

\bibitem[Murthy et al.(2010)]{Murthy2010} Murthy, J., Henry, 
R.~C., \& Sujatha, N.~V.\ 2010, \apj, 724, 1389 

\bibitem[Schlegel et al.(1998)]{Schlegel1998} Schlegel, D.~J., 
Finkbeiner, D.~P., \& Davis, M.\ 1998, \apj, 500, 525 

\bibitem[Seon et al.(2011)]{Seon2011} Seon, K.-I., Edelstein, 
J., Korpela, E., et al.\ 2011, \apjs, 196, 15

\bibitem[Sternberg(1989)]{Sternberg1989} Sternberg, A.\ 1989, \apj, 
347, 863

\bibitem[Sujatha et al.(2010)]{Sujatha2010} Sujatha, N.~V., Murthy, 
J., Suresh, R., Henry, R.~C., \& Bianchi, L.\ 2010, \apj, 723, 1549 

\bibitem[Witt 
\& Lillie(1973)]{Witt1973} Witt, A.~N., \& Lillie, C.~F.\ 1973, \aap, 25, 397 

\bibitem[Wright(1992)]{Wright1992} Wright, E.~L.\ 1992, \apj, 391, 
34


\end{thebibliography}
\end{document}